\documentclass[a4paper,apj]{emulateapj}
\usepackage{natbib,amsmath}
\shorttitle{NIR Spectroscopy of nearby Seyfert Galaxies}
\shortauthors{Terao et al.}

\begin{document}

\title{Near-infrared spectroscopy of nearby Seyfert galaxies:\\
   Is there evidence for the shock excitation in narrow-line regions?}

\author{
   K. Terao\altaffilmark{1},
   T. Nagao\altaffilmark{2},
   T. Hashimoto\altaffilmark{3},
   K. Yanagisawa\altaffilmark{4},
   K. Matsuoka\altaffilmark{5},
   Y. Toba\altaffilmark{2,6},
   H. Ikeda\altaffilmark{7}, and
   Y. Taniguchi\altaffilmark{8}}

\altaffiltext{1}{Graduate School of Science and Engineering, Ehime
   University, Bunkyo-cho 2-5, Matsuyama, Ehime 790-8577, Japan}
\altaffiltext{2}{Research Center for Space and Cosmic Evolution, Ehime
   University, Bunkyo-cho 2-5, Matsuyama, Ehime 790-8577, Japan}
\altaffiltext{3}{National Tsing Hua University, No. 101, Section 2, 
  Kuang-Fu Road, Hsinchu, 30013, Taiwan}
\altaffiltext{4}{Okayama Astrophysical Observatory, National Astronomical
   Observatory of Japan, Honjo 3037-5, Kamogata-cho, Asaguchi, 
   Okayama 719-0232, Japan}
\altaffiltext{5}{Department of Astronomy, Kyoto University,
   Kitashirakawa-Oiwake-cho, Sakyo-ku, Kyoto 606-8502, Japan}
\altaffiltext{6}{Institute of Astronomy and Astrophysics, Academia Sinica,
   PO Box 23-141, Taipei 10617, Taiwan}
\altaffiltext{7}{National Astronomical Observatory of Japan, 
   Osawa 2-21-1, Mitaka, Tokyo, 181-8588, Japan}
\altaffiltext{8}{The Open University of Japan, Wakaba 2-11, Mihama-ku, 
   Chiba 261-8586, Japan}
\email{terao@cosmos.phys.sci.ehime-u.ac.jp}

\begin{abstract}

One of the unsettled, important problems in active galactic nuclei (AGNs) 
is the major ionization mechanism  of gas clouds in their narrow-line regions (NLRs).
In order to investigate this issue, we present our $J$-band spectroscopic observations 
for a sample of 26 nearby Seyfert galaxies.
In our study, we use the flux ratio of the following two forbidden emission lines, 
[Fe~{\sc ii}]1.257$\mu$m and [P~{\sc ii}]1.188$\mu$m because it is known that
this ratio is sensitive to the ionization mechanism.
We obtain the [Fe~{\sc ii}]/[P~{\sc ii}] flux ratio or its lower limit for 19 objects. 
In addition to our data, we compile this flux ratio (or its lower limit) of 
23 nearby Seyfert galaxies from the literature. 
Based on the collected data, we find  that three Seyfert galaxies show very 
large lower limits of [Fe~{\sc ii}]/[P~{\sc ii}] flux ratios ($\gtrsim$10);
NGC 2782, 5005, and Mrk 463. It is thus suggested that the
contribution of the fast shock in the gas excitation is significantly large  for them.
However, more than half of Seyfert galaxies in our sample show moderate
[Fe~{\sc ii}]/[P~{\sc ii}] flux ratios ($\sim$2), that is consistent to 
the pure photoionization by power-law ionizing continuum emission. 
We also find that the [Fe~{\sc ii}]/[P~{\sc ii}] flux ratio  shows no clear correlation 
with the radio loudness, suggesting that the radio jet is not the primary origin of 
shocks in NLRs of Seyfert galaxies. 
\end{abstract}


\section{Introduction}

For these two decades, it has been observationally suggested that 
the evolution of galaxies is intimately related to the growth of supermassive 
black holes (SMBHs) in galactic nuclei
based on various scaling relations between the mass
of SMBHs ($M_{\rm BH}$) and properties of their host galaxies
(e.g., \citealt{1998AJ....115.2285M,2003ApJ...589L..21M,2005Natur.433..604D};
see, for reviews, \citealt{1995ARA&A..33..581K,2013ARA&A..51..511K}).
It is also known that a significant fraction of galaxies shows the nuclear activity 
that is driven by the gas accretion onto a SMBH \citep{1984ARA&A..22..471R,1997ApJ...487..568H}.
Such active galactic nuclei (AGNs) emit huge radiative energy. In addition to the
strong radiation, AGNs sometimes show strong outflows in various forms that are
recognized as the radio jet \citep[e.g.,][]{1980ApJ...240..429W,2006AJ....132..546G}, 
broad absorption-line (BAL) feature \citep[e.g.,][]{1991ApJ...373...23W,2009ApJ...692..758G},
ultra-fast outflow (UFO; e.g., \citealt{2006AN....327.1012C,2010A&A...521A..57T}), and so on. 
The released radiative and mechanical energies are expected to give significant 
impacts on various properties of the inter-stellar medium (ISM) in host galaxies, 
that may lead to the termination of the star-forming activity. This ``AGN feedback'' 
possibly takes an important role in the galaxy-SMBH coevolution, and therefore 
extensive observational efforts have been made to understand the AGN feedback 
phenomena (see \citealt{2012ARA&A..50..455F} for a review).

For studying the AGN feedback, one interesting approach is focusing on the
narrow-line region (NLR) in AGNs. This is because (1) the typical spatial scale of
the NLR is $\sim 10^{1-4}$ pc, covering the spatial scale of host
galaxies, and (2) the ionized gas clouds in the NLR radiate a variety of emission
lines that enable us to diagnose physical and chemical properties of the ISM
in NLRs. It is widely believed that the dominant ionization mechanism of ionized
gas clouds in NLRs is the photoionization by ionizing photons from the AGN 
central engine \citep[e.g.,][]{1996A&A...312..365B,1997A&A...323...31K,
2004ApJS..153...75G}. However it has been sometimes claimed that the fast 
shocks may also contribute significantly to the ionization in NLRs of some AGNs 
\citep[e.g.,][]{1996AJ....112...81K,1998ApJ...495..680B,1999ApJ...513L.115W,
2007ApJ...666..794F}. 
The discrimination of ionization mechanisms of gas clouds in the NLR is 
important in terms of the AGN feedback, because it is closely related to 
the question either radiative or mechanical heating is important for 
the gas in NLRs. Unfortunately, emission-line diagnostics in the rest-frame 
optical and ultraviolet range are not powerful to 
discriminate the ionization mechanism of NLR gas clouds, since the 
photoionization by the power-law spectral energy distribution (SED) and the fast 
shock result in very similar emission-line flux ratios \citep[e.g.,][]
{1995ApJ...455..468D,1996ApJS..102..161D,2008ApJS..178...20A}.

A powerful method to distinguish the ionization mechanism of NLR gas clouds
was proposed by \citet{2001A&A...369L...5O}. They reported that the flux ratio of 
two forbidden emission lines seen in near-infrared $J$-band, 
[P~{\sc ii}]1.188$\mu$m and [Fe~{\sc ii}]1.257$\mu$m, is useful to distinguish 
the ionization mechanism. The reason is as follows. These two transitions have 
similar critical densities ($n_{\rm cr}$ = 5.3$\times$10$^4$ cm$^{-3}$ for 
[P~{\sc ii}]1.188$\mu$m and $n_{\rm cr}$ = 3.5$\times$10$^4$ cm$^{-3}$ for 
[Fe~{\sc ii}]1.257$\mu$m at 10,000 K;
\citealt{2013Sci...342.1346K}), and the ionization 
potentials of the related ions are both below 13.6 eV that is the hydrogen
ionization potential (7.9 eV for [Fe~{\sc ii}] and 10.5 eV for [P~{\sc ii}]). 
Therefore these two lines should arise at similar locations in NLRs; more 
specifically, they are both mainly from the partially-ionized gas
\citep[see, e.g.,][]{2000ApJ...528..186M}. Consequently their flux ratio should be 
almost proportional to the gas-phase abundance ratio of the phosphorus and 
iron. Here the iron is a well known refractory species, 
and thus most of iron is locked into dust grains usually. 
On the other hand, the depletion of phosphorus onto dust grains is only moderate
\citep[see, e.g.,][]{1991ApJ...377L..57C}. This means that the 
gas-phase elemental abundance ratio of the iron to the phosphorus in NLR gas 
clouds photoionized by the power-law ionizing radiation is significantly lower 
than the solar elemental abundance ratio due to the selective depletion of the 
iron. However dust grains are easily destroyed due to fast shocks, and thus 
the relative abundance of the gas-phase iron in shock-dominated NLRs should 
be much higher than that in photoionized gas clouds. Therefore, the flux ratio of 
[Fe~{\sc ii}]/[P~{\sc ii}] is expected to be much higher in shock-dominated 
clouds ($\gtrsim 20$) than in photoionized clouds ($\sim 2$). Observations of 
nearby AGNs show that the measured [Fe~{\sc ii}]/[P~{\sc ii}] flux ratio is 
actually in the range of $\sim$2--20 \citep{2009MNRAS.394.1148S,
2010MNRAS.404..166R,2011MNRAS.417.2752R,2014MNRAS.445..414S}, and 
the flux ratio varies also with the location in the NLR \citep[e.g.,][]{
2009MNRAS.394.1148S,2010MNRAS.404..166R,2014MNRAS.442..656R,
2015MNRAS.451.3587R,2011PASJ...63L...7H,2011MNRAS.417.2752R,
2014MNRAS.445..414S}. Note that this flux ratio is totally free from the dust 
reddening thanks to the narrow wavelength separation between the two lines, 
which makes this flux ratio a powerful diagnostic tool for investigating the 
ionization mechanism of NLRs.

Though the [Fe~{\sc ii}]/[P~{\sc ii}] flux ratio is an interesting tool to explore the
nature of the AGN feedback at the NLR scale, this emission-line flux ratio has not
been measured in many objects so far ($\sim$ 20 objects; e.g., 
\citealt{2001A&A...369L...5O,2006ApJ...645..148R,
2009ApJ...694.1379R,2006A&A...457...61R,2013MNRAS.430.2249R,
2007MNRAS.376..719J,2007MNRAS.374..385K,2011PASJ...63L...7H,
2011MNRAS.417.2752R,2014MNRAS.445..414S,2015MNRAS.451.3587R}). 
Therefore it is hard to carry 
out various statistical analyses on the [Fe~{\sc ii}]/[P~{\sc ii}] flux ratio currently. 
Also the small sample size does not allow us to investigate possible connections 
between the NLR ionization and AGN outflows such as the radio jet, BAL, and 
UFO, given the fact that the fraction of AGNs showing those outflow features is 
relatively low ($\sim$10\% for radio jets, $\sim$10\% for BALs, and 
$\sim$35\% for UFOs; see, e.g., \citealt{2010A&A...521A..57T}).

In this paper, we present the result of our $J$-band spectroscopic observations
of the central region in a number of nearby Seyfert galaxies aiming at investigating 
the [Fe~{\sc ii}]/[P~{\sc ii}] flux ratio and examining the ionization mechanism of 
gas clouds in the NLR. In Section 2, we describe the details of our observations 
and the data reduction procedure. In Section 3, we summarize the obtained results.
We discuss the implications of our results in Section 4, and then we give the
conclusion in Section 5. 

\section{Observations and Data Reduction}

The targets were selected based on the optical [O~{\sc iii}]$\lambda$5007 flux
\citep[given by][]{1992ApJS...79...49W,1997ApJS..112..315H,1999ApJ...522..157R}
for focusing on objects with relatively strong NLR emission lines. Near-infrared 
long-slit spectroscopic observations were carried out for 26 nearby Seyfert 
galaxies and associated standard stars (for correcting the telluric absorption 
features and the wavelength dependence of the sensitivity) from August 2010 
to April 2011 with ISLE \citep{2006SPIE.6269E..3QY,2008SPIE.7014E..37Y}, 
a near-infrared imager and spectrograph for the Cassegrain focus of the 188 cm 
telescope at Okayama Astrophysical observatory (OAO). 
The detector (a HAWAII 1k $\times$ 1k array) provides a 4$^{\prime}$.3 
$\times$ 4$^{\prime}$.3 field of view with 0$^{\prime\prime}$.25 pixel$^{-1}$ 
spatial sampling. The observations were performed in $J$-band 
(1.11$-$1.32$\mu$m) with using a $2^{\prime\prime}.0$-width longslit,
and the typical nodding width was $\sim30^{\prime\prime}$ ($\sim$120 pixels). 
The slit was oriented to PA=90$^\circ$ (E-W) for all targets.
The unit exposure time was 120 sec, and the total on-source integration times
for science targets was 18--288 min, depending on the target.
The achieved spectral resolution is $\sim 1300$, that is measured from the
width of observed OH airglow emission lines.
The typical seeing size was $\sim 1^{\prime\prime}.0-2^{\prime\prime}.0$, that
is inferred from the spatial extension of standard stars on the 2 dimensional
spectra.
We summarize the redshifts, $J$-band magnitude, the AGN activity type, total exposure time,
date of observations for each target, and also the corresponding standard star,
in Table 1.

\begin{table*}[ht]
  \begin{center}
  \caption{Observation log}
    \begin{tabular}{lccccllc} \hline\hline
     Name & $z$\tablenotemark{a} & $m_{J}$\tablenotemark{b} & Type\tablenotemark{c} & Exposure time &  Date & Standard star\tablenotemark{d} & $m_{J}$\tablenotemark{e} \\ 
         &     & (mag)   &               &      (min)    &     &  & (mag)      \\ \hline
    NGC 1667 & 0.0152 & 9.92 & Sy 2 & 114 & 2011 January 13 & HIP 20507 (A2V) & 5.09 \\ 
    NGC 2273 & 0.0061 & 9.45 & Sy 2 & 94 & 2011 January 15, 16 & HIP 30060 (A2V) & 4.28\\ 
    NGC 2782 & 0.0085 & 9.79 & Sy 2 & 552 & 2011 January 13, 15 & HIP 45590 (A1V)& 5.76\\
     & & & & & 2011 March 24, 27 &  \\  
    NGC 3079 & 0.0037 & 8.44 & Sy 2 & 68 & 2011 January 12 & HIP 47006 (A2V) & 4.63 \\ 
    NGC 3982 & 0.0037 & 9.77 & Sy 1.9 & 18 & 2011 April 25 & HIP 62402 (A9V) & 5.45 \\ 
    NGC 4102 & 0.0028 & 8.76 & LINER & 110 & 2011 April 21, 28 & HIP 62402 (A9V)& 5.45 \\ 
    NGC 4169 & 0.0126 & 10.06 & Sy 2 & 68 & 2011 January 12 & HIP 60746 (A4V) & 4.80 \\ 
    NGC 4192 & $-$0.0005 & 7.82 & LINER & 48 & 2011 April 26 & HIP 59819 (A3V) & 5.17 \\ 
    NGC 4258 & 0.0015 & 6.38 & Sy 2 & 48 & 2011 April 23 & HIP 64906 (A0V) & 5.15 \\ 
    NGC 4388 & 0.0084 & 8.98 & Sy 2 & 36 & 2011 April 24 & HIP 59819 (A3V) & 5.17 \\ 
    NGC 4419 & $-$0.0009 & 8.78 & LINER\tablenotemark{f} & 60 & 2011 April 24 & HIP 59819 (A3V) & 5.17 \\ 
    NGC 4941 & 0.0037 & 9.09 & Sy 2 & 48 & 2011 April 24 & HIP 62983 (A2V) & 5.87 \\ 
    NGC 5005 & 0.0032 & 7.46 & LINER & 100 & 2011 April 23 & HIP 62641 (A3V) & 5.40 \\ 
    NGC 5194 & 0.0015 & 6.40 & Sy 2 & 168 & 2010 August 14$-$18 & HIP 66234 (A5V) & 4.53 \\ 
    NGC 5506 & 0.0062 & 9.71 & Sy 2 & 212 & 2011 April 23, 28 & HIP 68092 (A8V) & 5.45 \\ 
    NGC 6500 & 0.0100 & 10.13 & LINER & 208 & 2010 August 16$-$18 & HIP 91118 (A0V) & 5.63 \\
     & & & & & & HIP 87192 (A1V) & 5.99 \\ 
    NGC 6951 & 0.0048 & 8.31 & Sy 2 & 74 & 2010 August 12 & HIP 102253 (A8V) & 5.18 \\
    Mrk 3    & 0.0135 & 10.03 & Sy 2 & 56 & 2011 January 12 & HIP 29997 (A0V) & 4.97 \\
    Mrk 6    & 0.0188 & 11.08 & Sy 1.5 & 64 & 2011 January 13 & HIP 29997 (A0V) & 4.97 \\ 
    Mrk 34   & 0.0505 & 12.77 & Sy 2 & 26 & 2011 April 26 & HIP 52478 (A0V) & 5.80 \\
    Mrk 463  & 0.0504 & 12.16 & Sy 2 & 160 & 2011 April 24, 26, 28 & HIP 68276 (A0V) & 5.67 \\
    Mrk 477  & 0.0377 & 12.95 & Sy 2 & 84 & 2010 August 10, 12 & HIP 71573 (A1V) & 5.76 \\
    Mrk 509  & 0.0344 & 11.58 & Sy 1.5 & 166 & 2010 August 14, 15 & HIP 102618 (A1V) & 3.85 \\
    Mrk 766  & 0.0129 & 11.10 & Sy 1 & 84 & 2011 April 26 & HIP 60746 (A4V) & 4.80 \\
    Mrk 1073 & 0.0233 & 11.26 & Sy 2 & 288 & 2010 August 12, 14$-$18 & HIP 15648 (A3V) & 4.80 \\
    MCG +08-11-011 & 0.0205 & 10.49 & Sy 1.5 & 100 & 2011 January 12 & HIP 25143 (A3V) & 5.14 \\
     \hline
    \end{tabular}
  \end{center}
  \tablenotetext{1}{Redshift taken from the NASA/IPAC Extragalactic Database (NED).}
  \tablenotetext{2}{Apparent total $J$-band magnitude (in Vega), taken from 
    2MASS All-Sky Extended Source Catalog \citep{2006AJ....131.1163S}.}
  \tablenotetext{3}{Activity classification is taken from \citet{2010A&A...518A..10V} 
    unless a remark is given.} 
  \tablenotetext{4}{Spectral type of each standard star is given in the parenthesis.} 
  \tablenotetext{5}{Apparent $J$-band magnitude (in Vega) taken from 
    2MASS All-Sky Catalog of Point Sources \citep{2003yCat.2246....0C}.}
  \tablenotetext{6}{\citet{2014A&A...564A..67B}.}

\end{table*}

The data analysis was performed with using IRAF software 
\citep{1986SPIE..627..733T,1993ASPC...52..173T} by the standard manner; 
i.e., the flat fielding, background subtraction, stacking the 
individual spectra, spectral extraction from 2 dimensional spectra, wavelength 
calibration, correction for the atmospheric absorption and sensitivity function using
the spectra of standard stars, and finally the conversion of the spectra from
the observed frame to the rest frame. 
We adopted a $2^{\prime\prime}.0$ (8 pixels) aperture for the spectral extraction, by
taking account of the seeing size during the observations.
The magnitude and spectral type of standard stars were
collected from the 2MASS ({\it Two Micron All Sky Survey}) All-Sky
Catalog of Point Sources \citep{2003yCat.2246....0C} and the Hipparcos and
Tycho catalogues \citep{1997ESASP1200.....E}, respectively.
During the process of the correction for the atmospheric absorption and sensitivity 
function, a blackbody spectrum with the temperature corresponding to the
spectral type was assumed as the intrinsic spectrum.
The fluxes, central wavelengths, and FWHMs (full width at half maximums)
for the detected emission lines were
measured with an IRAF command {\tt splot}, assuming a single Gaussian for each 
emission line.
For undetected emission lines, we calculate the 3$\sigma$ flux upper limit based on
the rms of the spectra around the expected wavelength assuming a velocity width of
500 km s$^{-1}$ in FWHM, that is the typical value for detected emission lines in
our observations.

\section{Results}

\begin{figure*}[ht]
  \centering
  \epsscale{1.}
  \plotone{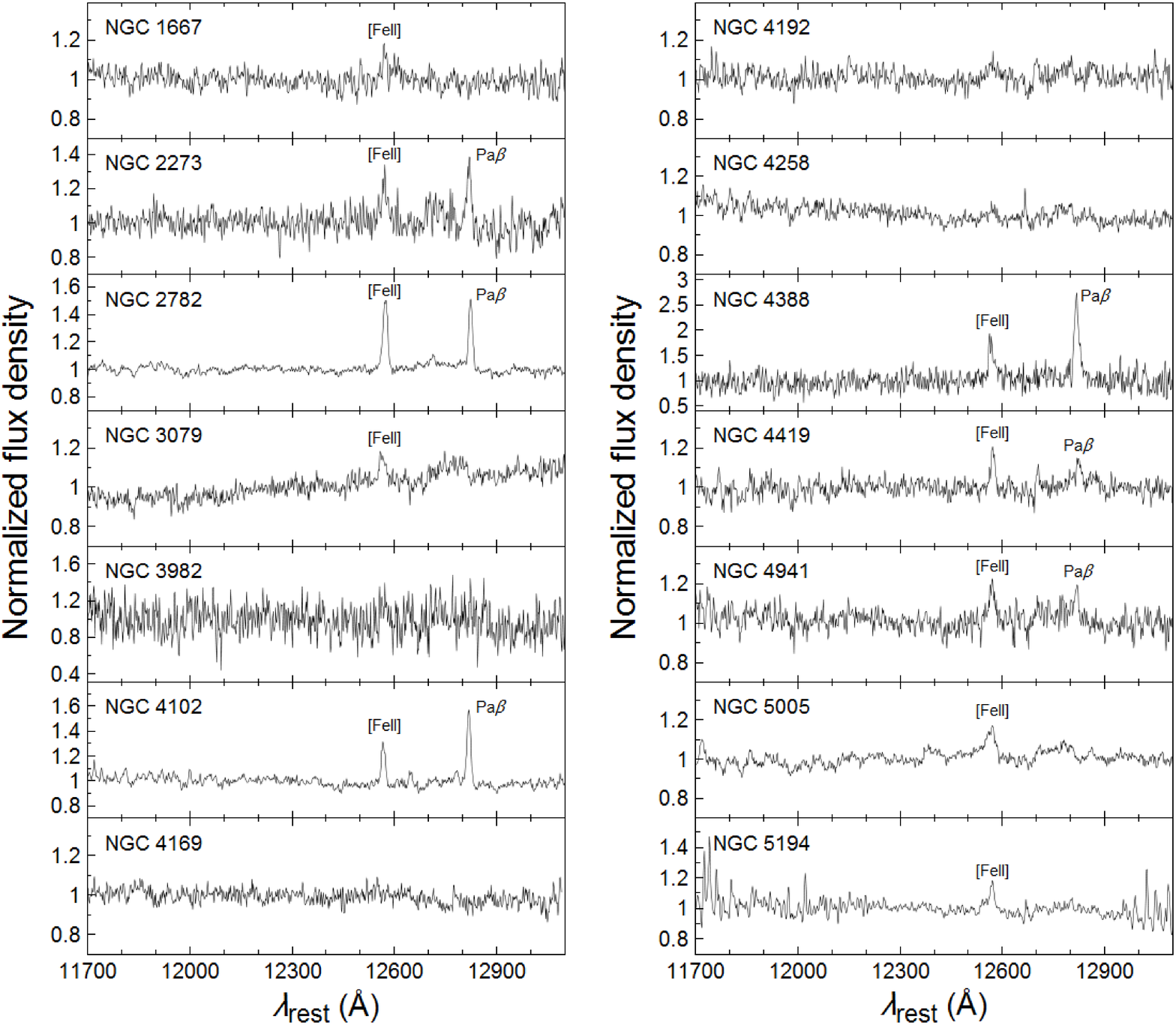}
  \caption{Rest-frame $J$-band spectra of our sample galaxies. Flux density is normalized to
 the 1.235$\mu$m continuum level.}
  \label{}
\end{figure*}

\setcounter{figure}{0}
\begin{figure*}[h]
  \centering
  \epsscale{1.}
  \plotone{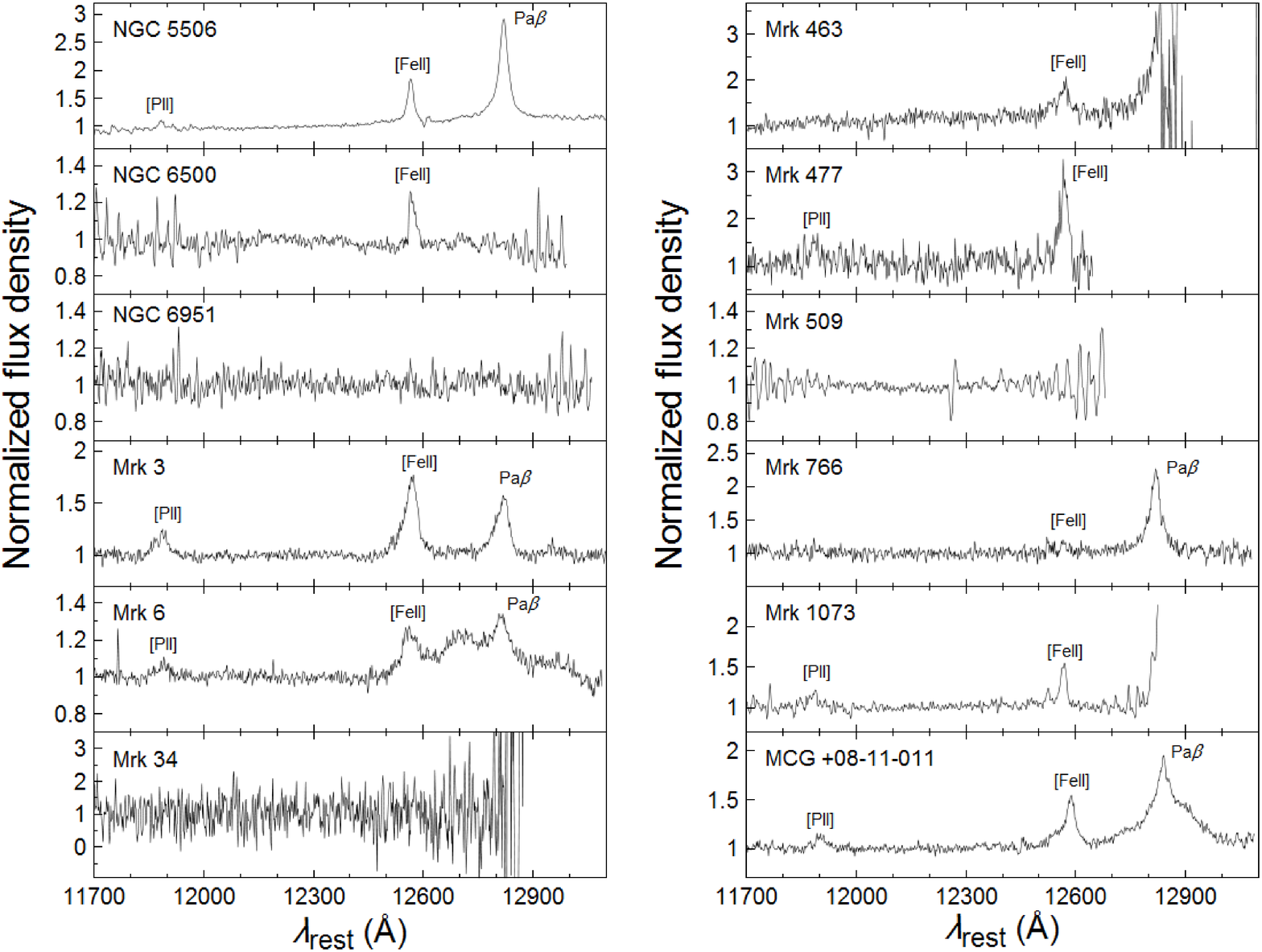}
  \caption{(Continued.)}
  \label{}
\end{figure*}

\begin{table*}
  \begin{center}
    \caption{Detected emission lines\tablenotemark{a}}
    \begin{tabular}{lcccccccc}
      \hline\hline
      \colhead{name} & \multicolumn{2}{c}{[P~{\sc ii}] 1.188$\mu$m} & &
      \multicolumn{2}{c}{[Fe~{\sc ii}] 1.257$\mu$m} & & \multicolumn{2}{c}{Pa$\beta$\tablenotemark{b} 1.282$\mu$m} \\
      \cline{2-3}\cline{5-6}\cline{8-9}
      & $\lambda_{\rm obs}$ & FWHM & & $\lambda_{\rm obs}$ & FWHM & & $\lambda_{\rm obs}$ & FWHM \\
      \hline
      NGC 1667 & $-$ & $-$ & & 1.2763 & 164 & & $-$ & $-$ \\
      NGC 2273 & $-$ & $-$ & & 1.2648 & 354 & & 1.2897 & 258 \\ 
      NGC 2782 & $-$ & $-$ & & 1.2681 & 300 & & 1.2933 & 200 \\
      NGC 3079 & $-$ & $-$ & & 1.2611 & 446 & & $-$ & $-$ \\
      NGC 3982 & $-$ & $-$ & & $-$ & $-$ & & $-$ & $-$ \\
      NGC 4102 & $-$ & $-$ & & 1.2603 & 229 & & 1.2855 & 175 \\
      NGC 4169 & $-$ & $-$ & & $-$ & $-$ & & $-$ & $-$ \\
      NGC 4192 & $-$ & $-$ & & $-$ & $-$ & & $-$ & $-$ \\
      NGC 4258 & $-$ & $-$ & & $-$ & $-$ & & $-$ & $-$ \\
      NGC 4388 & $-$ & $-$ & & 1.2566 & 511 & & 1.2926 & 324 \\
      NGC 4419 & $-$ & $-$ & & 1.2561 & 301 & & 1.2812 & 511 \\
      NGC 4941 & $-$ & $-$ & & 1.2616 & 475 & & 1.2865 & 233 \\
      NGC 5005 & $-$ & $-$ & & 1.2604 & 1251 & & $-$ & $-$ \\
      NGC 5194 & $-$ & $-$ & & 1.2587 & 548 & & $-$ & $-$ \\
      NGC 5506 & 1.1957 & 340 & & 1.2644 & 468 & & 1.2900 & 808 \\
      NGC 6500 & $-$ & $-$ & & 1.2696 & 474 & & $-$ & $-$ \\
      NGC 6951 & $-$ & $-$ & & $-$ & $-$ & & $-$ & $-$ \\
      Mrk 3 & 1.2045 & 1050 & & 1.2737 & 1091 & & 1.2989 & 1072 \\
      Mrk 6 & 1.2113 & 1017 & & 1.2797 & 1062 & & $-$ \tablenotemark{c} & $-$ \tablenotemark{c}\\
      Mrk 34 & $-$ & $-$ & & $-$ & $-$ & & $-$ \tablenotemark{d} & $-$ \tablenotemark{d} \\
      Mrk 463 & $-$ & $-$ & & 1.3199 & 1212 & & $-$ \tablenotemark{d} & $-$ \tablenotemark{d} \\
      Mrk 477 & 1.2333 & 480 & & 1.3042 & 691 & & $-$ \tablenotemark{d} & $-$ \tablenotemark{d} \\
      Mrk 509 & $-$ & $-$ & & $-$ & $-$ & & $-$ \tablenotemark{d} & $-$ \tablenotemark{d} \\
      Mrk 766 & $-$ & $-$ & & 1.2729 & 276 & & 1.2984 & 888 \\
      Mrk 1073 & 1.2160 & 605 & & 1.2859 & 470 & & $-$ \tablenotemark{d} & $-$ \tablenotemark{d} \\
      MCG +08-11-011 & 1.2127 & 812 & & 1.2821 & 957 & & $-$ \tablenotemark{c} & $-$ \tablenotemark{c} \\
    \hline
    \end{tabular}
  \end{center}
  \tablenotetext{1}{Observed central wavelength ($\mu$m)
    and FWHM (km s$^{-1}$) where the latter is corrected for the instrumental
    broadening.}
  \tablenotetext{2}{Narrow component.}
  \tablenotetext{3}{Not measured due to complex spectral features.}
  \tablenotetext{4}{Out of the wavelength coverage in our observations.}
\end{table*}

The fully reduced spectra of the 26 Seyfert galaxies are shown in Figure 1.
Here the flux scale is normalized by the flux density of the continuum emission
at $\lambda_{\rm rest}$ = 1.235$\mu$m. The [P~{\sc ii}] emission is significantly 
detected in 6 objects (NGC 5506, Mrk 3, Mrk 6, Mrk 477, Mrk 1073, and MCG 
+08-11-011) among the 26 targets, while 19 objects show the significant [Fe~{\sc ii}] 
emission. The Pa$\beta$ emission is also detected in 11 objects, but we do not 
measure the emission-line properties of Pa$\beta$ for Mrk 6 and MCG +08-11-011 
because of their very complex spectral profiles. The measured central wavelength 
and velocity width in FWHM for each detected emission line are given in Table 2, 
and the derived flux ratios of [Fe~{\sc ii}]/[P~{\sc ii}] and [Fe~{\sc ii}]/Pa$\beta$ 
are given in Table 3. The averages and standard deviations of velocity width of
detected emission lines are 717 $\pm$ 265 km s$^{-1}$, 596 $\pm$ 337 
km s$^{-1}$, and 497 $\pm$ 321 km s$^{-1}$ (for [P~{\sc ii}], [Fe~{\sc ii}], and
Pa$\beta$, respectively) in FWHM.

\begin{table*}
  \begin{center}
    \caption{Emission-line flux ratios of the targets}
    \begin{tabular}{lcc}
      \hline\hline
      Name & [Fe~{\sc ii}]/[P~{\sc ii}] & [Fe~{\sc ii}]/Pa$\beta$ \\
      \hline
      NGC 1667 & $> 3.174$ & $> 3.064$ \\
      NGC 2273 & $> 5.406$ & 0.876$\pm$0.086 \\
      NGC 2782 & $> 15.88$ & 1.218$\pm$0.019 \\
      NGC 3079 & $> 4.481$ & $> 4.048$ \\
      NGC 4102 & $> 7.427$ & 0.627$\pm$0.021 \\
      NGC 4388 & $> 5.896$ & 0.614$\pm$0.038 \\
      NGC 4419 & $> 4.067$ & 1.242$\pm$0.118 \\
      NGC 4941 & $> 4.880$ & 1.525$\pm$0.155 \\
      NGC 5005 & $> 12.37$ & $> 12.31$ \\
      NGC 5194 & $> 3.512$ & $> 6.660$ \\
      NGC 5506 & 6.876$\pm$0.645 & 0.278$\pm$0.003 \\
      NGC 6500 & $> 4.349$ & $> 8.259$ \\
      Mrk 3 & 3.974$\pm$0.162 & 1.314$\pm$0.023 \\
      Mrk 6 & 3.508$\pm$0.245 & $-$ \tablenotemark{a} \\
      Mrk 463 & $> 17.46$ & $-$ \tablenotemark{b} \\
      Mrk 477 & 4.330$\pm$0.414 & $-$ \tablenotemark{b} \\
      Mrk 766 & $> 2.280$ & 0.071$\pm$0.007 \\
      Mrk 1073 & 2.460$\pm$0.168 & $-$ \tablenotemark{b} \\
      MCG +08-11-011 & 4.256$\pm$0.220 & $-$ \tablenotemark{a} \\
    \hline
    \end{tabular}
  \end{center}
  \tablecomments{3$\sigma$ lower limits.}
  \tablenotetext{1}{Not measured due to complex spectral features.}
  \tablenotetext{2}{Pa$\beta$ is out of the wavelength coverage in our observations.}
\end{table*}

In addition to the results obtained in our observations, we also collected flux ratios 
of [Fe~{\sc ii}]/[P~{\sc ii}] and [Fe~{\sc ii}]/Pa$\beta$ from the literature 
in order to expand the sample size to enable a better statistical 
analyses of the flux ratios.
The compiled flux ratios for 23 objects are summarized in Table 4.

\begin{table*}
  \begin{center}
    \caption{Emission-line flux ratios in the literature}
    \begin{tabular}{lccccc}
      \hline\hline
      Name & $z$\tablenotemark{a} & [Fe~{\sc ii}]/[P~{\sc ii}] & [Fe~{\sc ii}]/Pa$\beta$ & Type\tablenotemark{b} & Ref\tablenotemark{c} \\
      \hline
      NGC 34 & 0.0196 & 1.55$\pm$0.43 & 1.07$\pm$0.14 & Sy 2 & 1 \\ 
      NGC 1068 & 0.0038 & 1.33$\pm$0.06 & 0.46$\pm$0.02 & Sy 2 & 2 \\
      NGC 1275 & 0.0176 & 3.35$\pm$0.12 & 1.05$\pm$0.06 & Sy 1.5 & 1 \\
      NGC 2110 & 0.0078 & 5.61$\pm$1.07 & 5.45$\pm$0.32 & Sy 2 & 1 \\
      NGC 3227 & 0.0039 & 2.80$\pm$0.61 & 1.21$\pm$0.08 & Sy 1.5 & 1 \\
      NGC 4151 & 0.0033 & 2.98$\pm$0.21 & 0.54$\pm$0.03 &  Sy 1.5 & 1 \\
      NGC 5128 & 0.0018 & 3.0\tablenotemark{d} & 2.3\tablenotemark{d} & FR I\tablenotemark{e} & 3 \\
      NGC 5929 & 0.0083 & 4.3$\pm$1.1 & $-$ \tablenotemark{f} & LINER & 4 \\
      NGC 7212 & 0.0266 & 3.64$\pm$0.42 & 0.86$\pm$0.05 & Sy 2 & 5 \\ 
      NGC 7465 & 0.0065 & $> 3.42$\tablenotemark{g} & 1.28$\pm$0.34 & LINER & 5 \\
      NGC 7469 & 0.0163 & 1.76$\pm$0.27 & 0.39$\pm$0.03 & Sy 1.5 & 1 \\ 
      NGC 7674 & 0.0289 & 1.90$\pm$0.19 & 1.04$\pm$0.11 & Sy 2 & 1 \\ 
      Mrk 34 & 0.0505 & 3.2$\pm$1.0 & $-$ \tablenotemark{f} & Sy 2 & 4 \\
      Mrk 78 & 0.0372 & 1.82$\pm$0.39 & 0.55$\pm$0.11 & Sy 2 & 6 \\ 
      Mrk 79 & 0.0222 & 1.55$\pm$0.67 & 0.17$\pm$0.05 & Sy 1 & 7 \\ 
      Mrk 334 & 0.0219 & 2.17$\pm$0.47 & 0.58$\pm$0.03 & Sy 1.8 & 1 \\
      Mrk 348 & 0.0150 & 4.14$\pm$0.67 & 0.99$\pm$0.10 & Sy 2 & 5 \\ 
      Mrk 573 & 0.0172 & 5.00$\pm$1.07 & 1.08$\pm$0.11 & Sy 2 & 5 \\
      Mrk 766 & 0.0129 & 2.65$\pm$0.16 & 0.38$\pm$0.03 & Sy 1 & 8 \\
      Mrk 1066 & 0.0120 & 2.74$\pm$0.24 & 0.79$\pm$0.03 & Sy 2 & 5 \\ 
      Mrk 1157 & 0.0152 & 2.63$\pm$0.31 & 0.82$\pm$0.07 & Sy 2 & 9 \\ 
      Ark 564 & 0.0247 & 2.07$\pm$0.40 & 0.09$\pm$0.01 & LINER & 1 \\
      ESO 428$-$G014 & 0.0057 & 1.99$\pm$0.11 & 0.67$\pm$0.02 & Sy 2 & 1 \\ 
    \hline
    \end{tabular}
  \end{center}
  \tablenotetext{1}{Redshift taken from the NASA/IPAC Extragalactic Database (NED).}
  \tablenotetext{2}{Activity classification is taken from \citet{2010A&A...518A..10V} 
    unless a remark is given.}
  \tablenotetext{3}{References: (1) \citet{2006A&A...457...61R}; 
    (2) \citet{2011PASJ...63L...7H}; 
    (3) \citet{2007MNRAS.374..385K}; (4) \citet{2007MNRAS.376..719J}; 
    (5) \citet{2009ApJ...694.1379R}; (6) \citet{2006ApJ...645..148R};
    (7) \citet{2013MNRAS.430.2249R}; (8) \citet{2014MNRAS.445..414S}; 
    (9) \citet{2011MNRAS.417.2752R}.}
  \tablenotetext{4}{The error of the flux ratio is not given in the literature.}
  \tablenotetext{5}{\citet{2003ApJ...593..169H}.}
  \tablenotetext{6}{Not available in the literature.}
  \tablenotetext{7}{3$\sigma$ lower limit.}
\end{table*}

\begin{figure}[htbp]
  \epsscale{1.2}
  \plotone{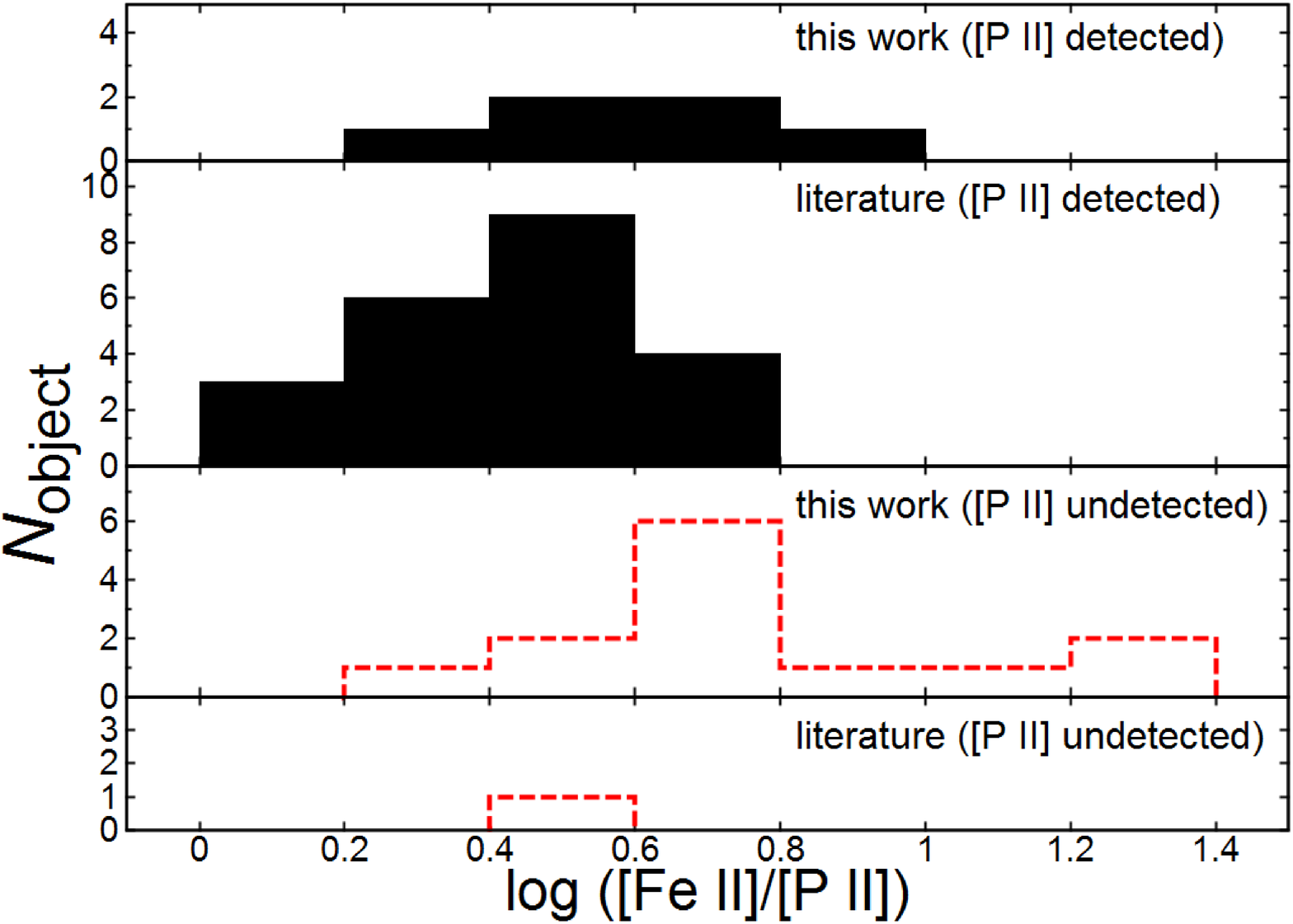}
  \caption{Histograms for our sample and the compiled one. 
    Filled-black histograms denote [P~{\sc ii}]-detected data
    while red-dashed histograms denote [P~{\sc ii}]-undetected data. The data 
    of [P~{\sc ii}]-detected objects in our OAO sample, [P~{\sc ii}]-undetected
    objects in our OAO sample, [P~{\sc ii}]-detected objects in the literature, and
    [P~{\sc ii}]-undetected objects in the literature, are shown from the top to the 
    bottom panel. For the [P~{\sc ii}]-detected objects, 3$\sigma$ lower limit of
    the [Fe~{\sc ii}]/[P~{\sc ii}] flux ratio is shown.
    }
  \label{}
\end{figure}

\begin{figure}[htbp]
  \epsscale{1.2}
  \plotone{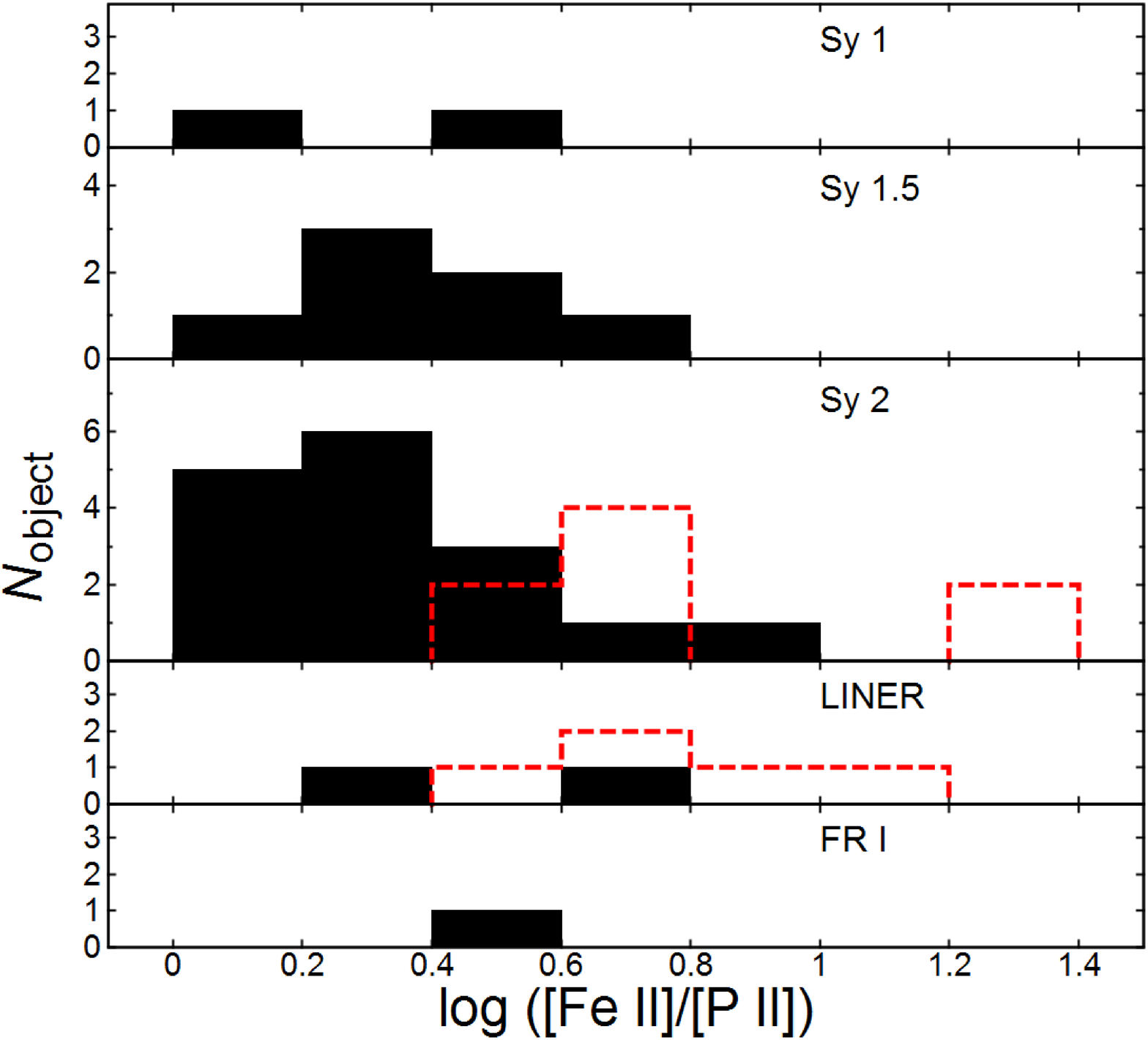}
  \caption{Same as Figure 2 but for each AGN type. 
    The type 1 Seyfert galaxies, type 1.5 Seyfert galaxies (including type 1.8
    Seyfert galaxies), type 2 Seyfert galaxies, LINERs, and FR I radio galaxies,
    are shown from the top to the bottom panel, respectively. The filled-black and
    red-dashed histograms are the same as in Figure 2.
    }
  \label{}
\end{figure}

Figure 2 shows the histograms of the measured [Fe~{\sc ii}]/[P~{\sc ii}] flux ratio or its
lower limit of the objects observed in our OAO observations and also compiled from 
the literature. The averages and standard deviations of the [Fe~{\sc ii}]/[P~{\sc ii}]
flux ratio for [P~{\sc ii}]-detected objects in our OAO sample and in the sample from
the literature are 4.24 $\pm$ 1.47 and 2.83 $\pm$ 1.15, respectively.
Note that Mrk 766 is included in both our OAO sample and also the sample 
compiled from the literature, and both of the two independent data of Mrk 766 are 
shown in Figure 2. Specifically, \citet{2014MNRAS.445..414S} reported that 
[Fe~{\sc ii}]/[P~{\sc ii}] = 2.654$\pm$0.164, that is consistent with our result
([Fe~{\sc ii}]/[P~{\sc ii}] $>$ 2.483). Hereinafter we adopt the flux ratio reported
by \citet{2014MNRAS.445..414S} for Mrk 766. Mrk 34 is also included in our 
sample and in the sample from the literature; we adopt the flux ratio reported by
\citet{2007MNRAS.376..719J} since our OAO observation did not detect any
emission lines (see Figure 1).
Our OAO observations show the presence of many nearby AGNs with
lower limits of [Fe~{\sc ii}]/[P~{\sc ii}] flux ratios, that have not been often 
reported in the literature.
This may be partly because the lower limit of the [Fe~{\sc ii}]/[P~{\sc ii}]
flux ratio for [P~{\sc ii}]-undetected objects has been unreported in the literature.

Figure 3 shows the histograms of the measured [Fe~{\sc ii}]/[P~{\sc ii}] flux ratio or its
lower limit for each AGN type. The averages and standard deviations of the 
[Fe~{\sc ii}]/[P~{\sc ii}] flux ratio for [P~{\sc ii}]-detected un-obscured (i.e., type 1 and
type 1.5) Seyfert galaxies (9 objects) and obscured (i.e., type 2) Seyfert galaxies
(16 objects) are  2.78 $\pm$ 0.82 and 3.32 $\pm$ 1.53, respectively.
Though there may be a tendency that objects showing a large [Fe~{\sc ii}]/[P~{\sc ii}] 
flux ratio are in the Seyfert 2 and LINER samples, we cannot conclude it statistically
because of small sample sizes in this work.
Note that a Fanaroff-Riley type I (FR I) galaxy, NGC 5128, shows a moderate value 
of the [Fe~{\sc ii}]/[P~{\sc ii}] flux ratio ($\sim$3.0; see Table 4), despite the presence
of a radio jet. We will discuss the implication for the relation between the radio jet
and the ionization mechanism of the NLR gas clouds inferred from the measured
[Fe~{\sc ii}]/[P~{\sc ii}] flux ratio in Section 4.2.

\section{Discussion}

\subsection{Emission-line flux ratios of the targets}

As shown in the previous section, we measured the flux ratio of
[Fe~{\sc ii}]/[P~{\sc ii}] for 6 objects with [P~{\sc ii}] detection in our OAO sample 
and collected 22 objects from the literature, 
accordingly for a total of 28 [P~{\sc ii}]-detected objects.
The average and standard deviation of this flux ratio for 
the 28 [P~{\sc ii}]-detected objects are 3.13 $\pm$ 1.33, and the median value is 2.88 
(see also Tables 3 and 4). This is roughly consistent with the prediction of 
photoionization models, that expect $\sim$2 for the [Fe~{\sc ii}]/[P~{\sc ii}] ratio 
\citep[e.g.,][]{2001A&A...369L...5O}. 
On the other hand, some objects show a much higher ratio of [Fe~{\sc ii}]/[P~{\sc ii}],
reaching at $\sim$10 or even more. As demonstrated by \citet[][]{2001A&A...369L...5O},
such a high [Fe~{\sc ii}]/[P~{\sc ii}] ratio suggests a significant contribution of fast 
shocks in the ionization of NLR gas clouds. It is therefore concluded that shock-dominated NLRs
actually exist, at least in some NLRs of AGNs in the nearby Universe.

Some previous long-slit and integral-field unit (IFU) observations show the spatial
variation of the [Fe~{\sc ii}]/[P~{\sc ii}] flux ratio in some nearby Seyfert galaxies, in
the sense that the [Fe~{\sc ii}]/[P~{\sc ii}] flux ratio increases as a function of the
distance from the nucleus \citep[e.g.,][]{2009MNRAS.394.1148S,
2010MNRAS.404..166R,2011PASJ...63L...7H}. The flux ratio reaches up to $\sim$10
at the outer part of NLRs in NGC 4151 \citep[][]{2009MNRAS.394.1148S} and Mrk 
1066 \citep[][]{2010MNRAS.404..166R}, suggesting that the shock contribution could 
be much more significant at the outer part than the inner part of NLRs. This trend is a 
natural consequence of the geometrical dilution of ionizing photons from the nucleus
(see also, e.g., \citealt{2007MNRAS.376..719J}). 
However, the surface emissivity of emission lines at the outer part of NLRs is much 
lower than the inner part of NLRs generally, and thus the integrated NLR spectrum of 
AGNs does not necessarily show shock-dominated line-flux ratios even if the outer 
part of NLRs shows such a high [Fe~{\sc ii}]/[P~{\sc ii}] flux ratio. Our OAO/ISLE 
spectra are extracted at the nuclear 2$^{\prime\prime}$.0 region, and the extracted 
spectra should not be dominated by the outer part of the NLR in the target AGNs. For
examining whether our OAO targets with a high flux ratio of [Fe~{\sc ii}]/[P~{\sc ii}] 
are affected by such a possible aperture effect or not, we show the relation between 
the measured [Fe~{\sc ii}]/[P~{\sc ii}] flux ratio and the physical scale covered by the 
aperture size (= 2$^{\prime\prime}$.0) at the target redshift in Figure 4. 
Though a positive correlation between the measured [Fe~{\sc ii}]/[P~{\sc ii}] flux ratio 
and the physical scale covered by the adopted aperture is expected if the high 
[Fe~{\sc ii}]/[P~{\sc ii}] flux ratio seen in our sample is caused by the significant
contribution of the outer part of NLRs, Figure 4 does not show such a significant
positive correlation. This suggests that the high [Fe~{\sc ii}]/[P~{\sc ii}] flux ratio seen 
in some objects in our OAO sample is not due to the geometrically-diluted ionizing 
photons at the outer part of NLRs.

It has been reported that the enhancement of the [Fe~{\sc ii}]/Pa~$\beta$ flux ratio is
seen in shock-excited regions, not only the [Fe~{\sc ii}]/[P~{\sc ii}] flux ratio (e.g.,
\citealt{1998ApJS..114...59L,2000ApJ...528..186M,2009ApJ...694.1379R,2015PKAS...30..145K}), 
that is due mostly to the destruction of dust grains by fast shocks.
Figure 5 shows the clear positive correlation between the [Fe~{\sc ii}]/[P~{\sc ii}] and
[Fe~{\sc ii}]/Pa~$\beta$ flux ratios, i.e., objects showing a large [Fe~{\sc ii}]/[P~{\sc ii}]
ratio also show a large [Fe~{\sc ii}]/Pa~$\beta$ ratio. This positive correlation has 
been reported by \citet[][]{2009ApJ...694.1379R} for a smaller sample (6 Seyfert
galaxies), but now the correlation is shown for a much larger sample in this work. 
This supports the interpretation that a large [Fe~{\sc ii}]/[P~{\sc ii}] ratio seen in NLRs 
of some AGNs is caused by fast shocks.

Note that, at high ionization parameter where Compton heating becomes 
significant, the photoionization may also destroy dust grains, resulting in higher 
[Fe~{\sc ii}]/[P~{\sc ii}] ratios without fast shocks. Such objects should be 
distinguished by checking the relative strength of coronal forbidden lines.
In order to examine this possibility, we investigate the relative strength of coronal 
forbidden lines in our samples. We find the [Fe~{\sc vii}]$\lambda$6087/[O~{\sc iii}]$\lambda$5007 
flux ratio for 21 objects (20 [P~{\sc ii}]-detected and 1 [P~{\sc ii}]-undetected objects) 
from the literature \citep{2000AJ....119.2605N,2003AJ....125.1729N,2006A&A...456..953B}.
Note that this flux ratio is very sensitive to the ionization parameter, 
in the sense that a higher [Fe~{\sc vii}]$\lambda$6087/[O~{\sc iii}]$\lambda$5007 
flux ratio is achieved by a higher ionization parameter 
(see, e.g., Figure 13 in \citealt{2001PASJ...53..629N}).
We also focus on a mid-infrared coronal line of [Ne~{\sc v}]14.32$\mu$m, and compile 
the [Ne~{\sc v}]14.32$\mu$m/[Ne~{\sc ii}]12.81$\mu$m flux ratio for 24 objects 
(17 [P~{\sc ii}]-detected and 7 [P~{\sc ii}]-undetected objects) from the literature 
\citep{2010ApJ...725.2270P,2010ApJ...716.1151W,2011ApJS..195...17W,
2011ApJ...730...28P,2011ApJ...740...94D}. 
Figures~\ref{coronal_Fe} and \ref{coronal_Ne} show 
the relation between the [Fe~{\sc ii}]/[P~{\sc ii}] flux ratio and the relative strength of 
two coronal forbidden lines. Though we expect a positive correlation in these figures if 
high-ionization gas clouds without fast shocks result in high [Fe~{\sc ii}]/[P~{\sc ii}] 
flux ratios, we do not see such a trend in both figures. These results further support our 
conclusion that a large [Fe~{\sc ii}]/[P~{\sc ii}] flux ratio seen in our samples is due to 
fast shocks, not to the photoionization with a very high ionization parameter.

\begin{figure}[tbp]
  \epsscale{1.2}
  \plotone{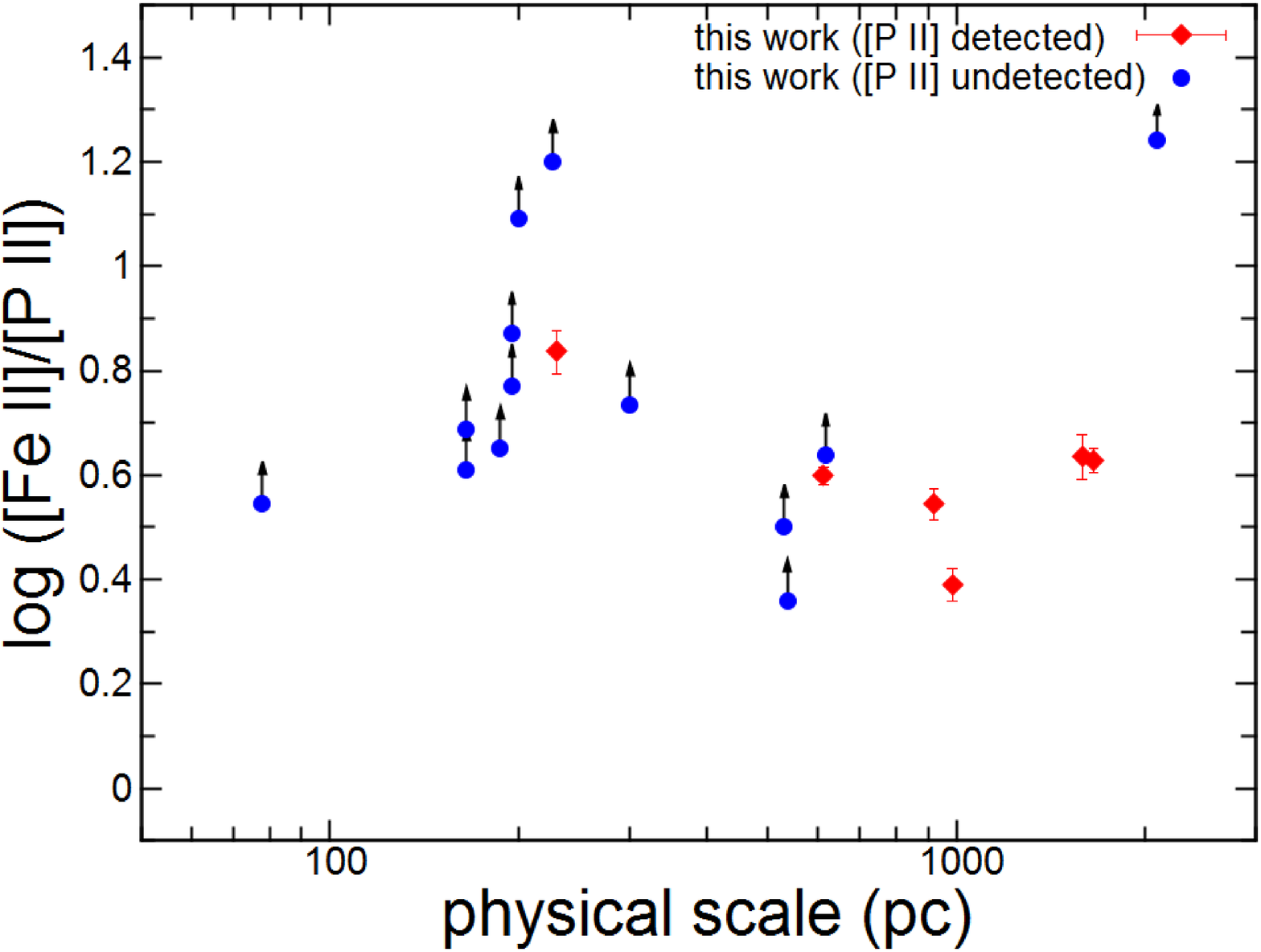}
  \caption{Relation between the measured [Fe~{\sc ii}]/[P~{\sc ii}] flux ratio and the 
  physical scale covered by the aperture size (= 2$^{\prime\prime}$.0) at the target 
  redshift. Red diamonds and blue circles denote [P~{\sc ii}]-detected and 
  [P~{\sc ii}]-undetected objects in our OAO sample, respectively. For the 
  [P~{\sc ii}]-undetected objects, the 3$\sigma$ lower limit of the [Fe~{\sc ii}]/[P~{\sc ii}] 
  flux ratio is shown.}
  \label{}
\end{figure}

\begin{figure}[tbp]
  \epsscale{1.2}
  \plotone{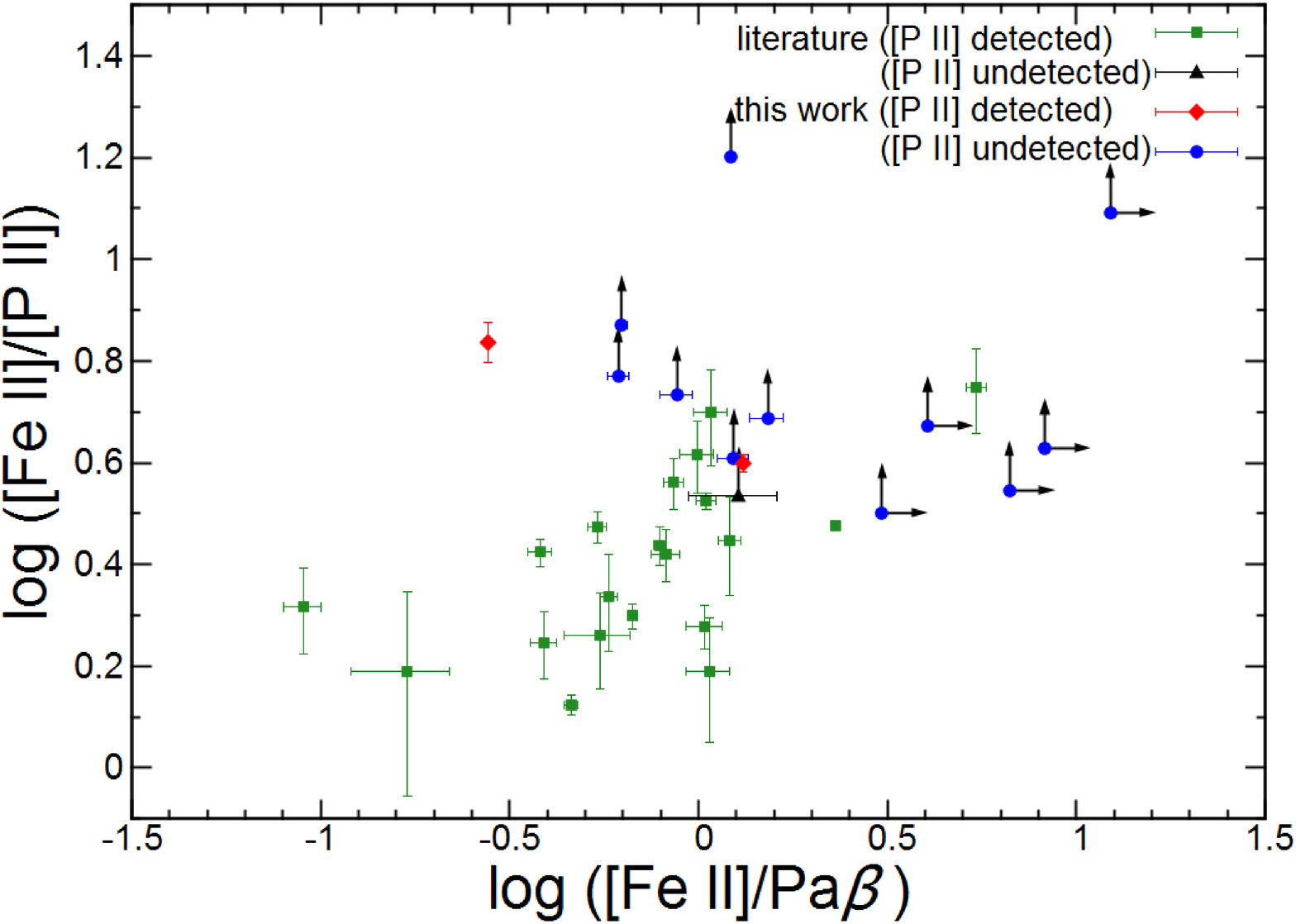}
  \caption{Relation between the [Fe~{\sc ii}]/[P~{\sc ii}] and [Fe~{\sc ii}]/Pa$\beta$ flux 
  ratios. Red diamonds and blue circles are the same as in Figure 4, while green 
  squares and black triangles denote [P~{\sc ii}]-detected and [P~{\sc ii}]-undetected 
  objects in the sample from the literature. Arrows denote 3$\sigma$ limits on the 
  emission-line flux ratio.
  }
  \label{}
\end{figure}

\begin{figure}[tbp]
  \epsscale{1.2}
  \plotone{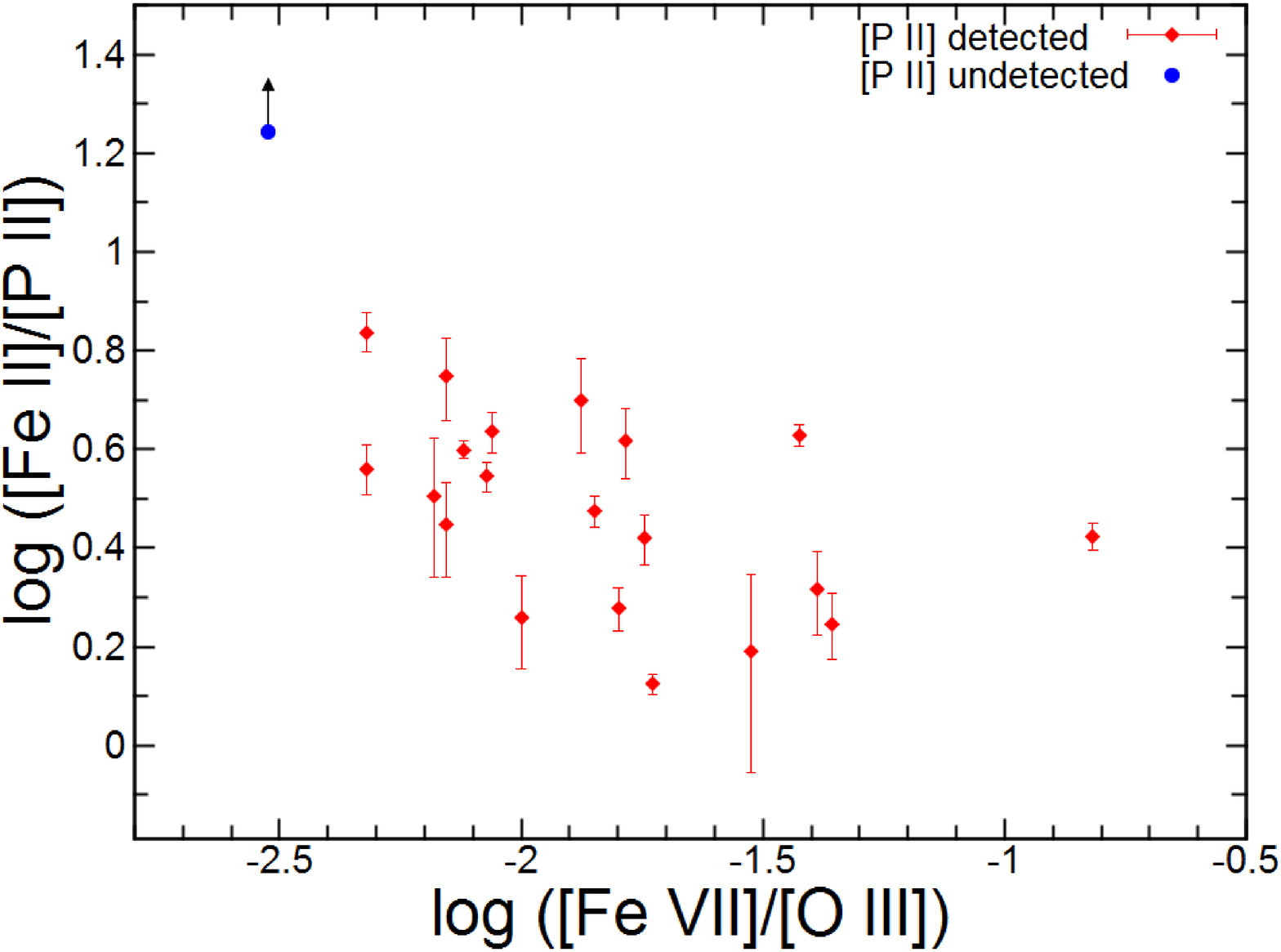}
  \caption{Relation between the [Fe~{\sc ii}]/[P~{\sc ii}] and 
    the [Fe~{\sc vii}]$\lambda$6087/[O~{\sc iii}]$\lambda$5007 flux
    ratios. Red diamonds and blue circles denote [P~{\sc ii}]-detected 
    and [P~{\sc ii}]-undetected objects, respectively. 
    Arrows denote 3$\sigma$ limits on the emission-line flux ratio.
  }
  \label{coronal_Fe}
\end{figure}

\begin{figure}[tbp]
  \epsscale{1.2}
  \plotone{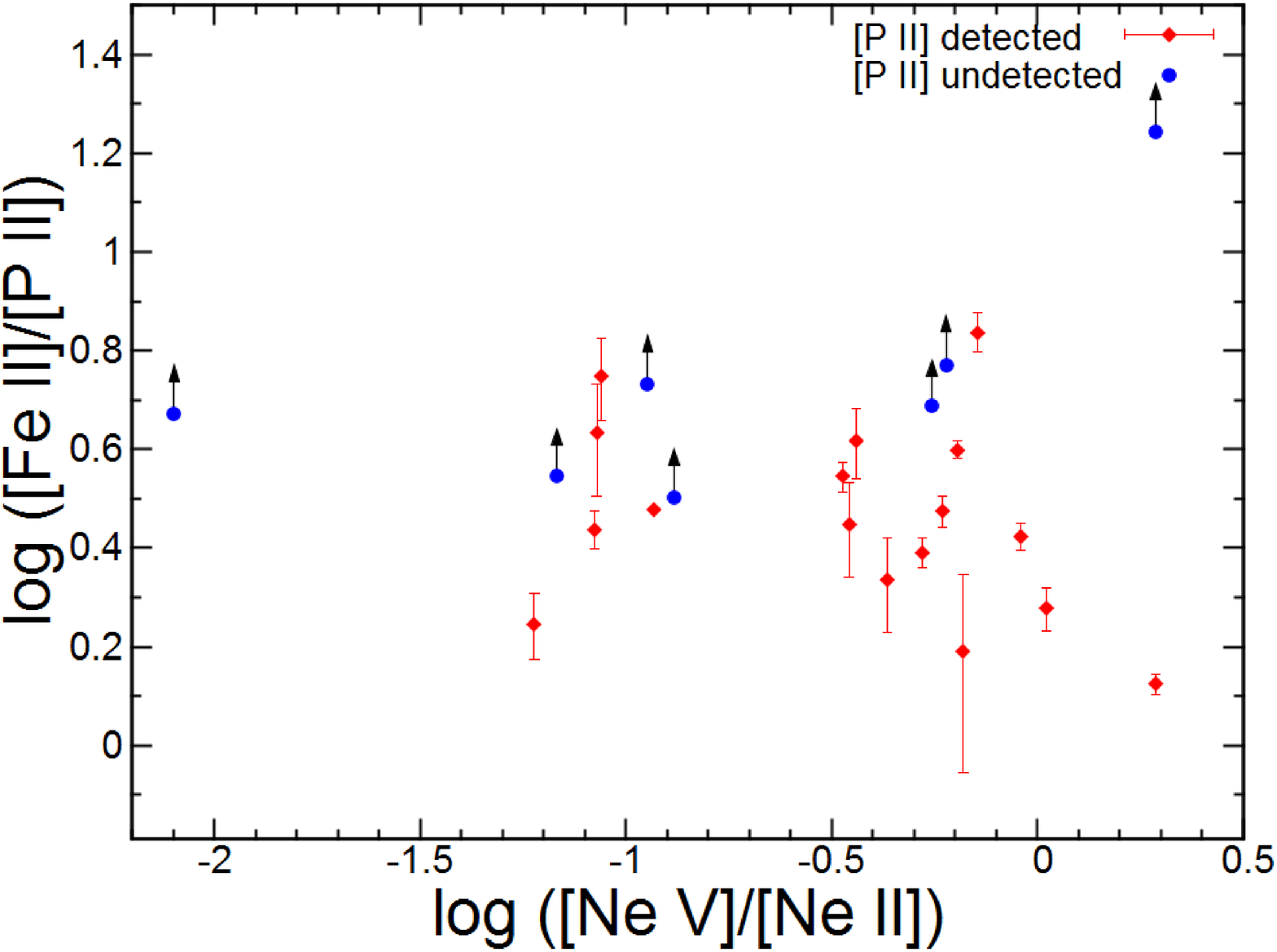}
  \caption{Relation between the [Fe~{\sc ii}]/[P~{\sc ii}] and 
    the [Ne~{\sc v}]14.32$\mu$m/[Ne~{\sc ii}]12.81$\mu$m flux 
    ratios. Symbols are the same as in Figure~\ref{coronal_Fe}.
  }
  \label{coronal_Ne}
\end{figure}

\subsection{Relation between the emission-line flux ratio and radio loudness}

\begin{figure}[tbp]
  \epsscale{1.2}
  \plotone{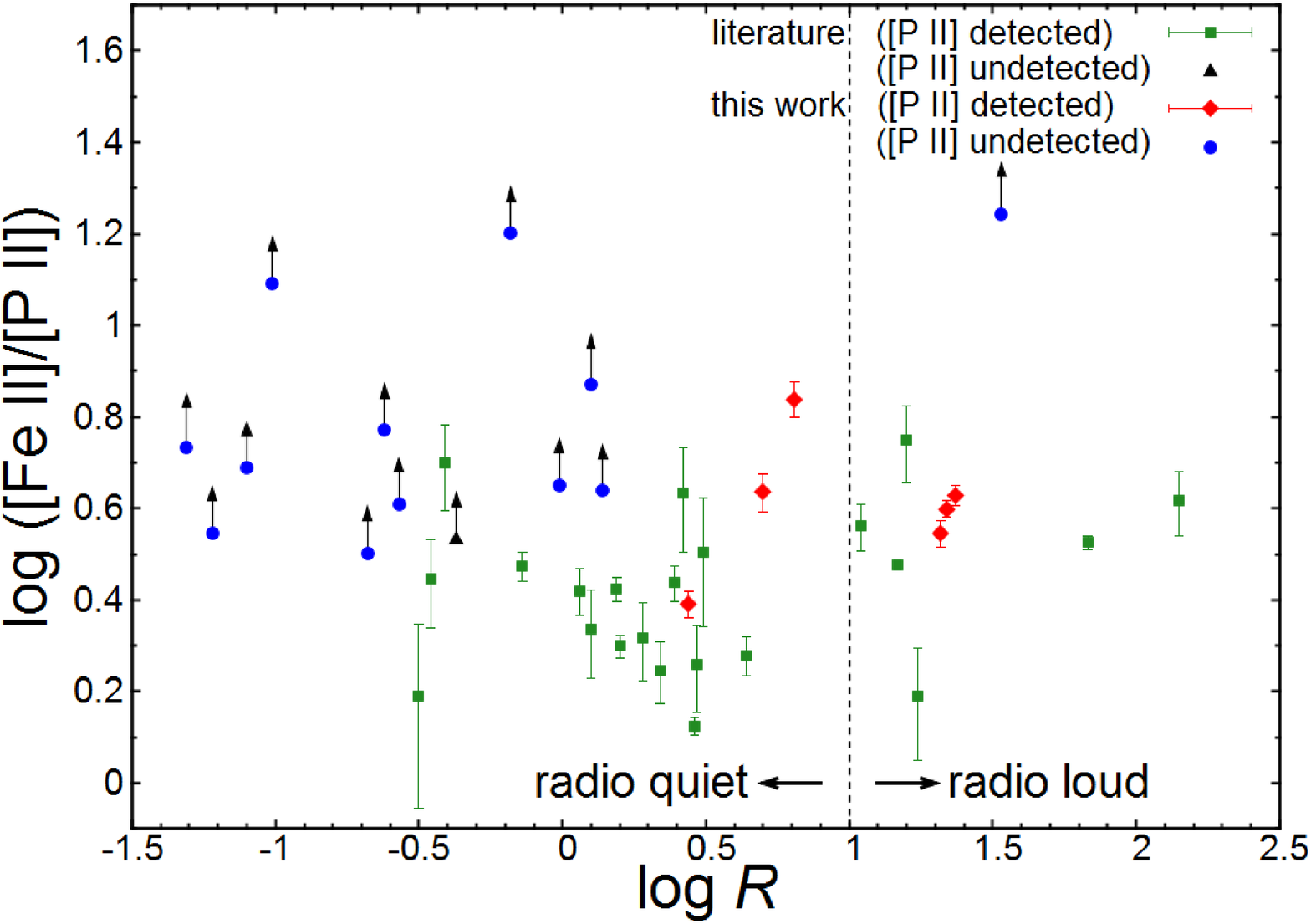}
  \caption{Relation between the [Fe~{\sc ii}]/[P~{\sc ii}] flux ratio and the radio-loudness
  ($R$). Symbols are the same as in Figure 5. Perpendicular dashed line shows the 
  threshold dividing the radio-loud and radio-quiet populations.}
  \label{radio_loudness}
\end{figure}

As described already, large [Fe~{\sc ii}]/[P~{\sc ii}] ratios seen in some AGNs are 
attributed by fast shocks in NLRs. The most simple idea for the origin of the fast 
shocks in NLRs is the effect of the AGN jet. Previous radio observations show 
that AGN jets exist not only in powerful radio galaxies; actually less-powerful radio 
jets are seen in many radio-quiet AGNs such as Seyfert galaxies, with much 
smaller spatial extension than seen in powerful radio galaxies
\citep[e.g.,][]{1993MNRAS.263..425M,1999ApJ...517L..81U,2001ApJ...559L..87N}. 
Therefore, such AGN jets including small-scale weak jets may cause fast shocks 
in NLRs and then the gas-phase iron abundance increases through the grain 
destruction or the sputtering process, resulting in a large [Fe~{\sc ii}]/[P~{\sc ii}] 
ratio observed in some AGNs.

For examining whether the observed large [Fe~{\sc ii}]/[P~{\sc ii}] ratio seen in 
some AGNs is actually attributed to the AGN jet or not, the most straightforward 
test is to investigate the relation between the [Fe~{\sc ii}]/[P~{\sc ii}] ratio and the 
power of the AGN jet. For quantifying the power of the AGN jet, we adopt the
radio loudness ($R$) that is defined as the ratio of the flux density of the radio
(5 GHz) to the optical (4400~\AA), i.e., 
\begin{equation}
     R = \frac{{\rm5\;GHz\; flux\; density}}{{\rm 4400\; \AA \;flux\;density}}
\end{equation}
\citep{1989AJ.....98.1195K}. 
In this definition, AGNs are classified as radio-loud and radio-quiet populations 
at the threshold of $R = 10$ \citep{1989AJ.....98.1195K}. We calculate the radio
loudness by compiling the optical and radio flux densities from previous 
observations, for both our OAO sample and the sample from the literature.
For the cases that multiple photometric data are available, we use the data
whose aperture size is as small as possible for avoiding any contributions from
the host galaxy. From the compiled photometric data, we derive the flux density
at 5 GHz and 4400~\AA \ by adopting the typical spectral index of Seyfert galaxies,
i.e., $-$0.5 (optical) and $-$0.7 (radio) respectively \citep{2015MNRAS.446..599S}.

Figure~\ref{radio_loudness} shows the [Fe~{\sc ii}]/[P~{\sc ii}] flux ratio as a 
function of the radio loudness, in the logarithmic scale. There is no apparent positive or negative
correlation between the [Fe~{\sc ii}]/[P~{\sc ii}] flux ratio and the radio loudness.
For the [P~{\sc ii}]-detected objects, there is no statistically significant difference
in the average value of the [Fe~{\sc ii}]/[P~{\sc ii}] flux ratio between the radio-loud
and radio-quiet populations; 3.67 $\pm$ 1.09 for 9 objects with $R>10$ (i.e., 
radio-loud, including NGC 5128 that is a FR I radio galaxy) while 2.87 $\pm$ 1.38 
for 19 objects with $R<10$ (i.e., radio-quiet).
We thus conclude that the [Fe~{\sc ii}]/[P~{\sc ii}] flux ratio is not determined 
primarily by the AGN jet. Note that this conclusion does not change even when
the [P~{\sc ii}]-undetected objects are taken into account, because there is only 
1 [P~{\sc ii}]-undetected (Mrk 463) in the radio-loud domain in our samples. This 
result strongly suggests that the radio jet is not the main origin of fast shocks in 
NLRs of AGNs.

\begin{table*}[ht]
  \begin{center}
    \caption{Various properties of our samples}
    \begin{tabular}{lccccccccccc} \hline\hline
    Name & [Fe~{\sc vii}]/[O~{\sc iii}]  &Ref& [Ne~{\sc v}]/[Ne~{\sc ii}] &Ref& $f_{\rm 5 GHz}$ & Ref & $f_{4400}$ & Ref & log $R$ & log $L_{\rm FIR}$ & log $\nu L_{\nu}$($K_{\rm s}$) \\ 
    & & & & & (mJy) & & (mJy) & & & ($L_{\odot}$) & ($L_{\odot}$)\\ 
    \hline
    \multicolumn{12}{c}{objects observed at OAO} \\
    \hline
    NGC 1667 & -- & & 0.131 & 4 & 5.0 & 9 & 24.1$\pm$7.08 & 19 & $-$0.68 & 10.55 & 10.38\\
    NGC 2273 & -- & & 0.113 & 5 & 2.4 & 10 & 49.4$\pm$9.44 & 19 & $-$1.31 & 10.00 & 10.05\\
    NGC 2782 & -- & & -- & & 44.05 & 11 & 66.9$\pm$9.91 & 19 & $-$0.18 & 9.89 & 9.65\\
    NGC 3079 & -- & & 0.008 & 5 & 120.15 & 11 & 124$\pm$17.1 & 19 & $-$0.01 & 10.47 & 10.12\\
    NGC 3982 & -- & & -- & & 1.63 & 11 & 82.7$\pm$13.1 & 19 & $-$1.71 & 9.80 & 9.59\\
    NGC 4102 & -- & & -- & & 91.60 & 11 & 72.7$\pm$10.8 & 19 & 0.10 & 10.48 & 9.61\\
    NGC 4169 & -- & & -- & & 0.59 & 11 & 23.0$\pm$4.14 & 19 & $-$1.59 & 10.29 & 10.32\\
    NGC 4192 & -- & & -- & & 6.88 & 11 & 222$\pm$21.4 & 19 & $-$1.51 & 9.49 & 10.09\\
    NGC 4258 & -- & & -- & & 2.43 & 11 & 819$\pm$121 & 19 & $-$2.53 & 9.46 & 10.01\\
    NGC 4388 & -- & & 0.600 & 5 & 18.46 & 11 & 76.1$\pm$5.82 & 19 & $-$0.62 & 9.86 & 9.86\\
    NGC 4419 & -- & & -- & & 16.20 & 11 & 60.5$\pm$4.62 & 19 & $-$0.57 & 9.64 & 9.82\\
    NGC 4941 & -- & & 0.554 & 5 & 4.92 & 11 & 61.6$\pm$14.5 & 19 & $-$1.10 & 8.93 & 9.63\\
    NGC 5005 & -- & & -- & & 18.78 & 11 & 191$\pm$30.4 & 19 & $-$1.01 & 10.25 & 10.51\\
    NGC 5194 & -- & & 0.068 & 5 & 6.04 & 11 & 994$\pm$147 & 19 & $-$1.22 & 9.71 & 10.06\\
    NGC 5506 & 0.0048 & 1 & 0.716 & 5 & 138.53 & 11 & 21.5$\pm$3.89 & 19 & 0.81 & 9.86 & 9.93\\
    NGC 6500 & -- & & -- & & 35.8 & 10 & 25.7$\pm$4.91 & 19 & 0.14 & 9.86 & 10.40\\
    NGC 6951 & -- & & -- & & 36$\pm$8 & 12 & 73.3$\pm$11.7 & 19 & $-$0.31 & 10.17 & 10.19\\
    Mrk 3 & 0.0076 & 1 & 0.640 & 5 & 355.59 & 13 & 16.2$\pm$2.75 & 19 & 1.34 & 10.33 & 10.47\\
    Mrk 6 & 0.0085 & 1 & 0.336 & 5 & 98.5 & 13 & 4.76$\pm$0.76 & 19 & 1.32 & 10.12 & 10.59\\
    Mrk 34 & 0.0066 & 1 & -- & & 7.91 & 11 & 2.56$\pm$0.71 & 20 & 0.49 & 10.73 & 10.49\\
    Mrk 463 & 0.0030 & 2 & 1.935 & 5 & 150.81 & 11 & 4.42$\pm$0.85 & 19 & 1.53 & 11.14 & 11.09\\
    Mrk 477 & 0.0087 & 1 & -- & & 24.28 & 11 & 4.89$\pm$1.35 & 20 & 0.70 & 10.75 & 10.07\\
    Mrk 509 & -- & & -- & & 7.87$\pm$0.41 & 14 & 7.76$\pm$2.14 & 20 & 0.01 & 11.07 & 11.22\\
    Mrk 766 & 0.1524 & 1 & 0.913 & 5 & 16.54 & 11 & 10.7$\pm$2.16 & 19 & 0.19 & 10.25 & 10.01\\
    Mrk 1073 & -- & & 0.525 & 6 & 38.58$\pm$1.19 & 14 & 14.1$\pm$4.49 & 19 & 0.44 & 11.13 & 10.42\\
    MCG +08-11-011 & 0.0376 & 1 & -- & & 100.90$\pm$3.57 & 14 & 4.26$\pm$0.86 & 21 & 1.37 & 11.13 & 11.26\\
    \hline
    \multicolumn{12}{c}{objects from the literature} \\
    \hline
    NGC 34 & -- & & -- & & 44.39$\pm$5.79 & 15 & 2.58$\pm$0.01 & 22 & 1.24 & 11.24 & 10.28\\
    NGC 1068 & 0.0187 & 1 & 1.941 & 5 & 1650.25 & 11 & 577$\pm$85.5 & 19 & 0.46 & 10.66 & 10.35\\
    NGC 1275 & -- & & -- & & 3100$\pm$160 & 16 & 45.4$\pm$5.78 & 19 & 1.83 & 10.71 & 10.89\\
    NGC 2110 & 0.0070 & 1 & 0.087 & 7 & 104.80$\pm$4.92 & 17 & 6.66$\pm$0.43 & 23 & 1.20 & 9.95 & 10.30\\
    NGC 3227 & 0.0070 & 1 & 0.348 & 5 & 33.96 & 11 & 98.5$\pm$16.7 & 19 & $-$0.46 & 9.85 & 10.08\\
    NGC 4151 & 0.0142 & 1 & 0.588 & 5 & 135.83 & 11 & 186$\pm$27.6 & 19 & $-$0.14 & 9.30 & 9.80\\
    NGC 5128 & -- & & 0.117 & 5 & 52662$\pm$2662 & 18 & 3540$\pm$756 & 19 & 1.17 & 9.65 & 10.04\\
    NGC 5929 & -- & & 0.085 & 5 & 28.95 & 11 & 10.92$\pm$0.29 & 24 & 0.42 & 10.35 & 10.05\\
    NGC 7212 & 0.0048 & 3 & -- & & 42.29 & 11 & 3.89$\pm$1.73 & 25 & 1.04 & 10.84 & 10.51\\
    NGC 7465 & -- & & -- & & 8.53$\pm$0.57 & 14 & 20.2$\pm$3.86 & 19 & $-$0.37 & 9.61 & 9.51\\
    NGC 7469 & 0.0440 & 1 & 0.060 & 5 & 59.73 & 11 & 27.4$\pm$3.77 & 19 & 0.34 & 11.27 & 10.58\\
    NGC 7674 & 0.0159 & 1 & 1.050 & 5 & 72.48 & 11 & 16.7$\pm$3.18 & 19 & 0.64 & 10.78 & 10.37\\
    Mrk 34 & 0.0066 & 1 & -- & & 7.91 & 11 & 2.56$\pm$0.71 & 20 & 0.49 & 10.73 & 10.49\\
    Mrk 78 & 0.0100 & 1 & -- & & 17.37$\pm$3.86 & 15 & 5.88$\pm$1.63 & 20 & 0.47 & 10.62 & 10.38\\
    Mrk 79 & 0.0299 & 1 & 0.660 & 5 & 6.09 & 11 & 19.3$\pm$3.69 & 19 & $-$0.50 & 10.64 & 10.80\\
    Mrk 334 & -- & & 0.433 & 4 & 9.65$\pm$3.86 & 15 & 7.68$\pm$1.38 & 19 & 0.10 & 10.82 & 10.36\\
    Mrk 348 & 0.0165 & 1 & 0.363 & 5 & 801.7 & 9 & 5.72$\pm$1.16 & 19 & 2.15 & 9.99 & 10.07\\
    Mrk 573 & 0.0133 & 1 & -- & & 5.64 & 11 & 14.4$\pm$2.13 & 19 & $-$0.41 & 9.99 & 10.05\\
    Mrk 766 & 0.1524 & 1 & 0.913 & 5 & 16.54 & 11 & 10.7$\pm$2.16 & 19 & 0.19 & 10.25 & 10.01\\
    Mrk 1066 & -- & & 0.084 & 8 & 36.67$\pm$3.86 & 15 & 14.9$\pm$2.21 & 19 & 0.39 & 10.62 & 9.97\\
    Mrk 1157 & 0.0180 & 1 & -- & & 13.65$\pm$0.45 & 14 & 11.8$\pm$2.14 & 19 & 0.06 & 10.17 & 10.08\\
    Ark 564 & 0.041 & 1 & -- & & 11.93$\pm$0.41 & 14 & 6.26$\pm$1.48 & 19 & 0.28 & 10.23 & 10.22\\
    ESO 428-G014 & -- & & -- & & 33.50$\pm$1.23 & 14 & 21.2$\pm$4.51 & 19 & 0.20 & 9.57 & 9.68\\
    \hline
    \end{tabular}
  \end{center}
  \tablecomments{References: (1) \citealt{2000AJ....119.2605N}; 
    (2) \citealt{2003AJ....125.1729N}; (3) \citealt{2006A&A...456..953B}; 
    (4) \citealt{2011ApJS..195...17W}; (5) \citealt{2010ApJ...725.2270P}; 
    (6) \citealt{2011ApJ...730...28P}; (7) \citealt{2010ApJ...716.1151W};
    (8) \citealt{2011ApJ...740...94D}; (9) \citealt{2006AJ....132..546G}; 
    (10) 1.4 GHz data from \citealt{2005A&A...435..521N};
    (11) FIRST catalog ({\it Faint Images of the Radio Sky at Twenty-cm}); 
    (12) \citealt{1975AJ.....80..771S}; (13) 4.89 GHz data from \citealt{1997A&AS..122..235L}; 
    (14) 1.4 GHz data from \citealt{1998AJ....115.1693C}; 
    (15) \citealt{1995ApJS...98..369B}; (16) \citealt{2006MNRAS.368.1500T}; 
    (17) 4.89 GHz data from \citealt{1984ApJ...285..439U}; 
    (18) 4.85 GHz data from \citealt{1994ApJS...90..173G}; 
    (19) \citealt{1991rc3..book.....D}, RC3 ({\it Third Reference Catalog of Bright Galaxies}), 
    photographic magnitude; 
    (20) 4680 \AA \ data from \citealt{2007ApJS..170...33P}; 
    (21) \citealt{1991rc3..book.....D} (RC3), total magnitude;
    (22) \citealt{2007AJ....133.2132S}; 
    (23) 4480 \AA \ data from \citealt{1983ApJS...52..341M};
    (24) SDSS, $g$-band (4860 \AA); 
    (25) \citealt{1968cgcg.bookR....Z}, CGCG ({\it Catalogue of Galaxies and Clusters of Galaxies}).}
\end{table*}

\subsection{Relation between the emission-line flux ratio and kinematics}

\begin{figure}[tbp]
  \epsscale{1.2}
  \plotone{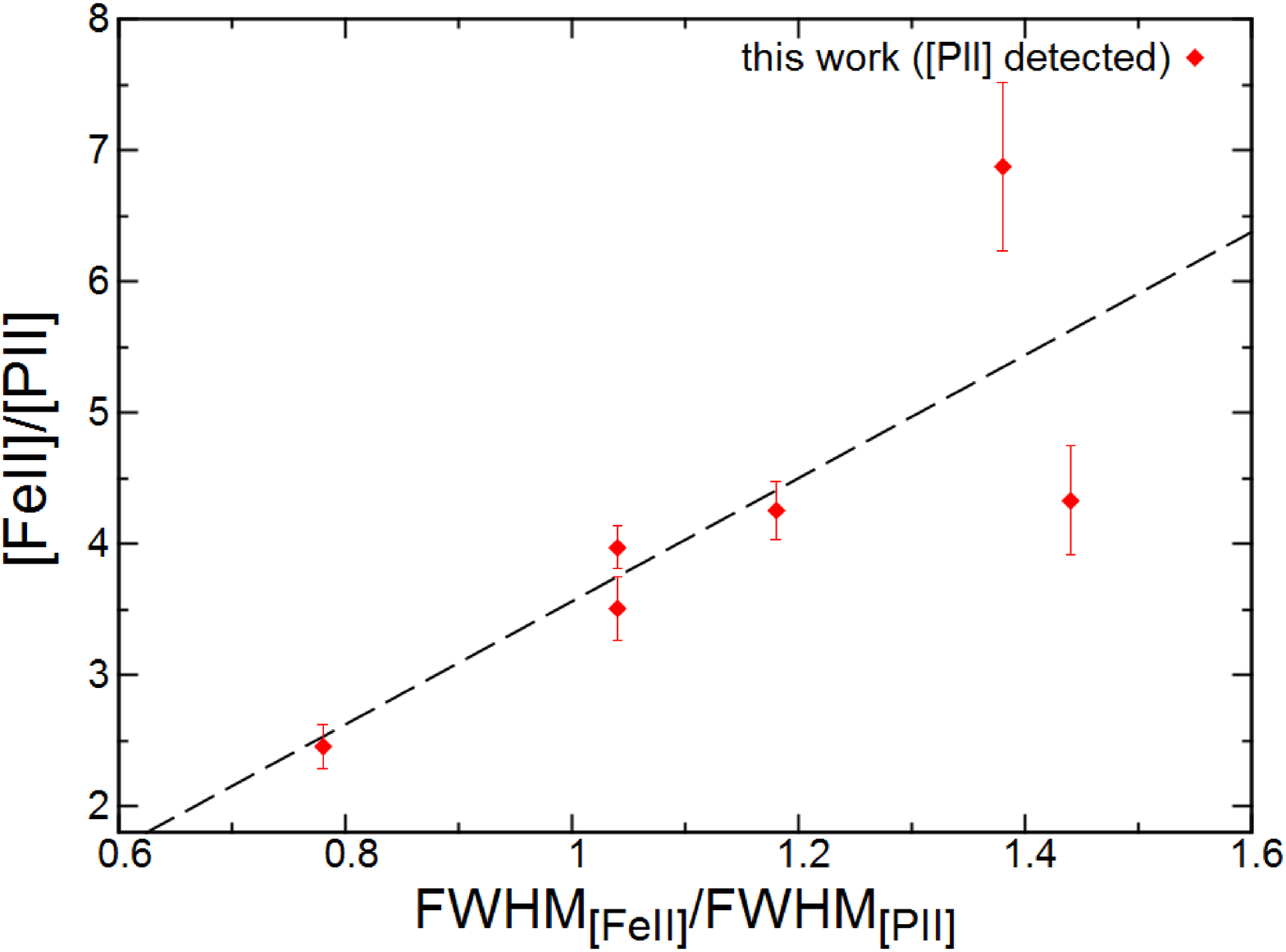}
  \caption{Relation between the [Fe~{\sc ii}]/[P~{\sc ii}] flux ratio and the FWHM ratio of 
    [Fe~{\sc ii}] and [P~{\sc ii}] for objects whose [Fe~{\sc ii}] and [P~{\sc ii}] are
    detected in our OAO run. Dashed line represents the linear-fit result.}
  \label{fwhm}
\end{figure}

As described in Section 1, the [Fe~{\sc ii}] and [P~{\sc ii}] emission lines arise at 
similar locations in the NLR due to the similarity of the critical density and ionization 
potential. Therefore, the velocity profiles of these two lines are expected to be 
similar in each other. However, NLR clouds affected by fast shocks may be 
kinematically disturbed. Therefore the velocity profile of [Fe~{\sc ii}] and [P~{\sc ii}] 
emission lines may tell us some hints about the contribution of fast shocks in NLRs.
Figure~\ref{fwhm} shows the relation between the [Fe~{\sc ii}]/[P~{\sc ii}] flux ratio 
and the FWHM ratio of [Fe~{\sc ii}] and [P~{\sc ii}], and we see a significant positive
correlation between the two quantities. Specifically, AGNs showing a larger  
[Fe~{\sc ii}]/[P~{\sc ii}] flux ratio tend to show wider [Fe~{\sc ii}] velocity profile with
respect to the [P~{\sc ii}] velocity profile. The linear fit to this correlation results in
the relation of 
\begin{equation}
\begin{split}
     &{\rm [Fe~II]/[P~II]} = \\
     &(4.690 \pm 1.866) \times \frac{\rm FWHM_{[Fe~II]}}{\rm FWHM_{[P~II]}} + (-1.128 \pm 2.174)
\end{split}
\end{equation}
(see Figure~\ref{fwhm}). Note that this positive correlation is not expected if all of 
the observed flux of the [Fe~{\sc ii}] and [P~{\sc ii}] emission lines comes from
NLR clouds excited by fast shocks, because in this case both [Fe~{\sc ii}] and 
[P~{\sc ii}] velocity profiles should be disturbed similarly by the shock and thus
the FWHM ratio is expected to be $\sim$1. The observed correlation shown in
Figure~\ref{fwhm} is explained by the mixture of the photoionized cloud and
shock-excited cloud in NLRs. More specifically, the [Fe~{\sc ii}] flux of AGNs with
a large [Fe~{\sc ii}]/[P~{\sc ii}] flux ratio is significantly attributed by fast shocks
while the [P~{\sc ii}] flux is largely attributed by photoionized clouds in the NLR. 
This picture is consistent to the fact that the observed [Fe~{\sc ii}]/[P~{\sc ii}] flux
ratio does not reach up to the value expected by pure shocks ($\sim$20).

\subsection{Origin of the fast shock in NLRs}

As already described, our results show that the radio jet is not the primary origin of 
fast shocks in NLRs, at least low-$z$ AGNs investigated in this work. Then, what is 
the origin of fast shocks in NLRs? Recently, powerful AGN-driven outflows are 
observed in some high-$z$ \citep[e.g.,][]{2012MNRAS.425L..66M,
2015A&A...580A.102C} and low-$z$ AGNs \citep[e.g.,][]{2011ApJ...729L..27R,
2014A&A...567A.125G,2015A&A...583A..99F}. 
The possible launching mechanisms of powerful AGN outflows are the radiation pressure 
from the UV photons and/or line-force \citep[e.g.,][]{1995ApJ...451..498M,1997A&A...327..909B,
2008ApJ...676..101P,2016PASJ...68...16N}, 
thermally driven \citep[e.g.,][]{1983ApJ...271...70B,2001ApJ...561..684K},
magnetocentrifugal force due to an accretion disk 
\citep[e.g.,][]{1982MNRAS.199..883B,1994ApJ...434..446K}.
Those outflows are characterized by
broad (FWHM $\gtrsim$ 1000 km s$^{-1}$) and/or blueshifted emission lines, mostly 
extending up to the kpc scale. The opening angle of those outflows is much wider than 
that of radio jets generally, and such powerful outflows are seen not only in radio-loud
AGNs. Therefore it is strongly suggested that the main driving source of those AGN 
outflows is not the radio jet. For example, \citet{2015A&A...583A..99F} reported the 
discovery of a molecular outflow spreading $\sim$1 kpc toward all directions from the 
nucleus of a type 1 Seyfert Mrk 231 (that is known also as an ultra-luminous infrared 
galaxy; ULIRG) based on CO(2-1) observations. The inferred mass-outflow rate and 
kinetic energy of the molecular outflow are $\dot{M} = 500 - 1000\ M_{\odot} {\rm yr}^{-1}$ 
and $\dot{E}_{\rm kin} = 7 - 10 \times 10^{43}$ erg s$^{-1}$ that corresponds to 1--2\% 
of the AGN bolometric luminosity. Similarly, \citet{2014A&A...562A..21C} reported the 
ratio of the kinetic energy of the outflow to the AGN bolometric luminosity in 14 low-$z$ 
luminous AGNs (including ULIRGs) is up to 5\%, also based on CO observations.
The value of this ratio is close to predicted value from theoretical model for explaining 
the $M_{BH} - \sigma$ relation in local galaxies \citep[e.g.,][]{1998A&A...331L...1S,
2010MNRAS.402.1516K}. 
Therefore, it is likely that powerful AGN-driven outflows can
give impacts on NLR clouds and consequently cause fast shocks in NLRs.

However, the physical origin of such powerful AGN-driven outflows in the galactic
scale is also unknown. At a smaller spatial scale in AGNs, powerful outflows are 
sometimes recognized as the BAL and UFO. Since BAL features are seen in 
rest-frame UV spectra and thus they are observed mainly for high-$z$ quasars
\citep[e.g.,][]{1991ApJ...373...23W}, we here focus on the UFO that are seen in
rest-frame hard X-ray spectra of nearby AGNs \citep[e.g.,][]{2010A&A...521A..57T}.
Specifically, the UFO is identified as a blue-shifted Fe K absorption line, whose
outflowing velocity reaches mildly relativistic values ($>$ 0.033c). 
Based on hydrodynamical simulations, \cite{2013ApJ...763L..18W} showed that 
the uncollimated UFO from the central engine of AGNs gives strong feedback effects
into the ISM in the host galaxy at the kpc scale, similar to well-collimated radio jets
\citep[see, e.g.,][]{2012ApJ...757..136W}.
Among our sample, 5 objects (NGC 4151, NGC 5506, Mrk 79, Mrk 509, and Mrk 766)
show the UFO feature in their X-ray spectra \citep{2010A&A...521A..57T,
2013MNRAS.430...60G}. These objects show relatively low [Fe~{\sc ii}]/[P~{\sc ii}] flux 
ratios that reach only up to $\sim$3, except for NGC 5506 (6.876$\pm$0.487) and 
Mrk 509 (unmeasured).
This may suggest that AGNs with a shock-excited NLR are not necessarily associated
with the UFO feature in their X-ray spectra. Here it should be noted that the UFO could 
be still important as a possible origin of fast shocks in NLRs, because AGNs without 
UFO features may possess powerful nuclear outflows that are not recognized as UFOs
due to various effects such as the viewing-angle effect and obscuration effect.
A larger sample of AGNs with the measurement of the [Fe~{\sc ii}]/[P~{\sc ii}] flux ratio
and hard X-ray spectra is needed to investigate possible link between the fast shock
in NLRs and the UFO phenomenon.

\begin{figure}[tbp]
  \epsscale{1.2}
  \plotone{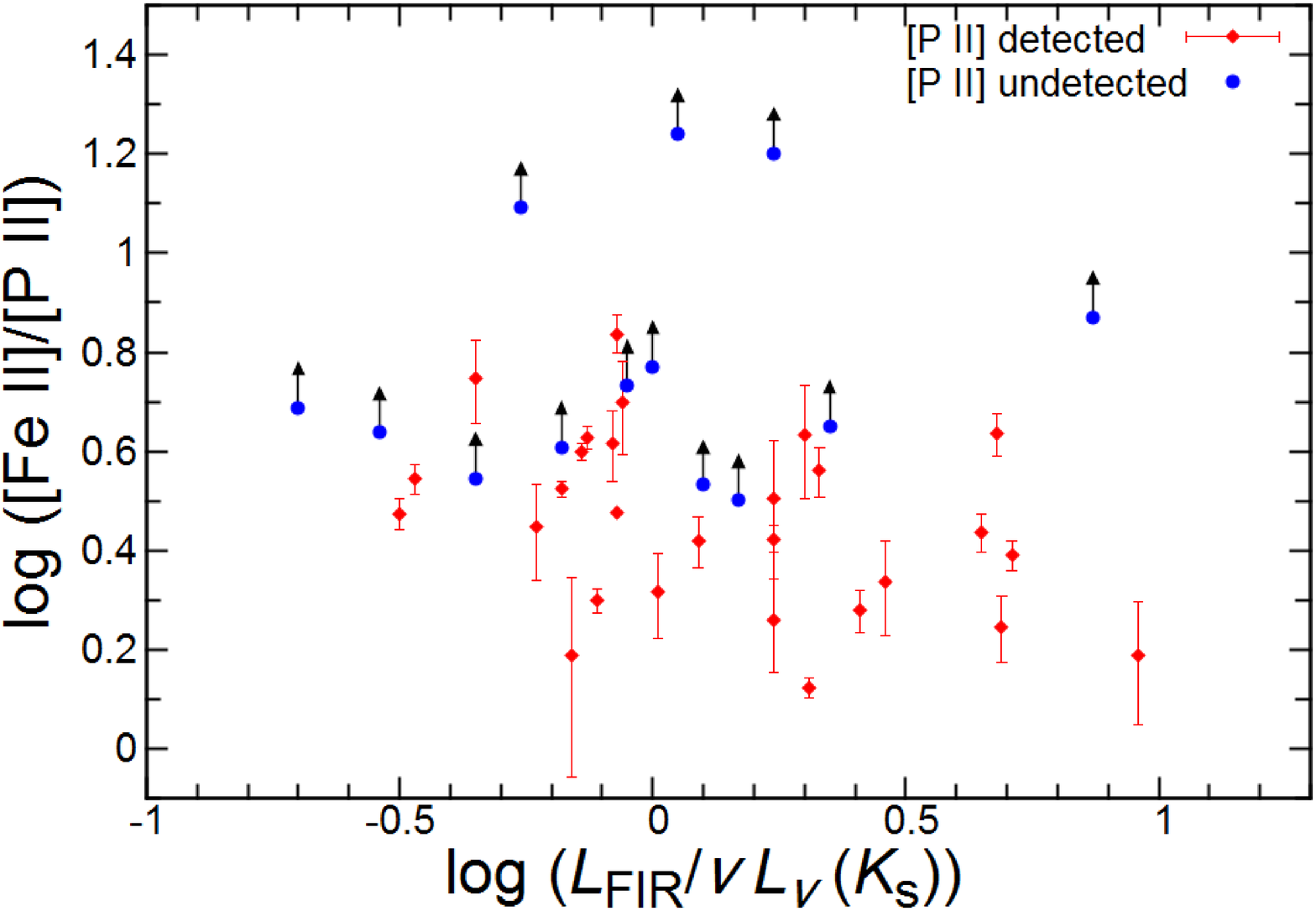}
  \caption{Relation between the [Fe~{\sc ii}]/[P~{\sc ii}] flux ratio and the 
  far-infrared luminosity normalized to the $K_{\rm s}$ luminosity. 
  Symbols are the same as in Figure~\ref{coronal_Fe}.}
  \label{L_FIR}
\end{figure}

Moreover, we discuss another possible candidate for the origin of 
fast shocks, that is the wind originated from the nuclear starburst
\citep[e.g.,][]{1990ApJS...74..833H,2010ApJ...721..505R,2014MNRAS.444.3894H,2016MNRAS.457.1257H}. 
In order to examine whether the starburst is the main origin of the fast shock in 
the NLR or not, we investigate two indicators of the starburst activity.
First, we focus on the far-infrared luminosity ($L_{\rm FIR}$), that is a tracer of the 
starburst activity in galaxies. We adopt the following definition of the far-infrared flux 
($F_{\rm FIR}$) to derive $L_{\rm FIR}$,
\begin{equation}
     F_{\rm FIR} = 1.26\times10^{-11}(2.58S_{60\mu\rm m}+S_{100\mu\rm m})
\end{equation}
where $S_{60\mu\rm m}$ and $S_{100\mu\rm m}$ are the {\it IRAS} flux densities in Jy at 60$\mu$m and 
100$\mu$m, respectively \citep{1989cgqo.book.....F,1996ARA&A..34..749S},
and $F_{\rm FIR}$ is given in unit of erg s$^{-1}$ cm$^{-2}$. 
Since the outflow due to the starburst is determined by both the 
starburst activity and the depth of the gravitational potential of galaxies, we investigate 
the far-infrared luminosity normalized by the $K_{\rm s}$-band luminosity ($\nu L_\nu$($K_{\rm s}$)), 
where the latter is derived by the total $K_{\rm s}$ magnitude taken from
the 2MASS All-Sky Extended Source Catalog \citep{2006AJ....131.1163S}.
Figure~\ref{L_FIR} shows the relation between the [Fe~{\sc ii}]/[P~{\sc ii}] flux ratio and 
$L_{\rm FIR}/\nu L_\nu$($K_{\rm s}$), and we find no significant positive correlation. 
This suggests that the large [Fe~{\sc ii}]/[P~{\sc ii}] flux ratio is not caused by 
the starburst activity in host galaxies of Seyfert galaxies.
Second, we checked the 3.3$\mu$m polycyclic aromatic hydrocarbon (PAH) feature 
which is indicator of the nuclear starburst 
\citep[e.g.,][]{2000ApJ...545..701I,2004ApJ...617..214I}.
Since the strong PAH emission is expected for starburst galaxies while the PAH is 
destroyed in pure AGNs, we can infer the relative strength of the starburst to 
the AGN through the PAH emission. 
We collected the 3.3$\mu$m PAH data from the literature 
\citep{2003MNRAS.340L..33R,2008ApJ...677..895W,2010PASJ...62.1509O,2014PASJ...66..110C}.
The averages and standard deviations of the [Fe~{\sc ii}]/[P~{\sc ii}] flux ratio
for the PAH detected objects (7 objects) and undetected (upper limit) objects (6 objects) 
in [P~{\sc ii}]-detected sample are 3.28$\pm$1.75 and 3.52$\pm$1.04, respectively. 
Since there is no systematic difference in the [Fe~{\sc ii}]/[P~{\sc ii}] flux ratio 
between the PAH-detected and PAH-undetected samples, the contribution of 
starburst-driven shocks to the [Fe~{\sc ii}]/[P~{\sc ii}] flux ratio appears 
to be negligible.
Therefore, based on the two tracers of the starburst, $L_{\rm FIR}$ and PAH, we conclude that 
the starburst is not the primary origin of fast shocks in NLRs.

\section{CONCLUSION}

In order to investigate how the fast shock contributes to the NLR ionization,
we have carried out near-infrared $J$-band spectroscopic observations of 26 nearby AGNs.
In our analysis, we use  the [Fe~{\sc ii}]/[P~{\sc ii}] flux ratio as a powerful diagnostics 
of the fast shock in the NLR. 
Among the 26 observed AGNs, the [P~{\sc ii}] emission is significantly detected
in 6 AGNs while the [Fe~{\sc ii}] emission is significantly detected in 19 AGNs. By 
adding the data from the literature, we gather the [Fe~{\sc ii}]/[P~{\sc ii}] flux ratio or
its lower limit for 23 nearby AGNs.
Based on this combined large sample of Seyfert galaxies, we obtain the following results
and conclusions.

\begin{enumerate}
\item[1.] We find that the [Fe~{\sc ii}]/[P~{\sc ii}] flux ratio in more than half of Seyfert
     galaxies in our sample are consistent to the
     prediction by the photoionization model ($\sim$2). However, the three Seyfert galaxies,
     NGC 2782, 5005, and Mrk 463 show very 
     large [Fe~{\sc ii}]/[P~{\sc ii}] flux ratios ($\gtrsim$10), suggesting a significant 
     contribution of the fast shock in the NLR excitation.
\item[2.] The positive correlation between the [Fe~{\sc ii}]/[P~{\sc ii}] flux ratio and the
     [Fe~{\sc ii}]/Pa~$\beta$ reported in the literature is confirmed in our large sample
      of Seyfert galaxies. This also supports the interpretation that the NLR in AGNs
     with a large [Fe~{\sc ii}]/[P~{\sc ii}] flux ratio is affected by fast shocks.
\item[3.] The positive correlation between the [Fe~{\sc ii}]/[P~{\sc ii}] flux ratio and the
     FWHM ratio of [Fe~{\sc ii}] and [P~{\sc ii}] is found in our sample. This is consistent
     with the interpretation that the observed [Fe~{\sc ii}] flux in AGNs with a large
     [Fe~{\sc ii}]/[P~{\sc ii}] flux ratio is significantly attributed by shock-excited clouds.
\item[4.] The [Fe~{\sc ii}]/[P~{\sc ii}] flux ratio in our sample shows no clear correlation 
     with the radio loudness
     nor the strength of the starburst, suggesting that the radio jet
     and the starburst are
     not the primary origin of the fast shocks in the NLR. 
\end{enumerate}


\acknowledgments
This paper is based on observations with the 188 cm telescope at the Okayama 
Astrophysical Observatory (OAO) operated by National Astronomical Observatory 
of Japan (NAOJ). We thank K. Matsubayashi, T. Hori, N. Araki, and the OAO 
staffs for their support during the observations. 
We would like to thank the 
anonymous referee for her/his useful comments and suggestions
that improved this paper very much.
This work was financially supported in part by the Japan Society for
the Promotion of Science (JSPS; TN: grant No. 25707010, 16H01101, and 16H03958;
YT: 23244031 and 16H02166), 
and also by the JGC-S Scholarship Foundation. KM is financially supported by
the JSPS through the JSPS Research Fellowship.
This research has made use of the NASA/IPAC Extragalactic Database (NED) 
which is operated by the Jet Propulsion Laboratory, California Institute of Technology, 
under contract with the National Aeronautics and Space Administration.
This publication makes use of data products from the Two Micron All Sky Survey, which 
is a joint project of the University of Massachusetts and the Infrared Processing and 
Analysis Center/California Institute of Technology, funded by the National Aeronautics 
and Space Administration and the National Science Foundation.
IRAF is distributed by the National Optical Astronomy Observatory, which is operated 
by the Association of Universities for Research in Astronomy (AURA) under a 
cooperative agreement with the National Science Foundation.


\bibliography{reference}

\end{document}